\documentclass[aps,eqsecnum,a4paper]{revtex4}
\usepackage{makeidx}
\usepackage{amsfonts}
\usepackage{amssymb}
\usepackage{mathrsfs}
\usepackage{ifthen}
\usepackage[mathscr]{euscript}
\usepackage{graphicx}
\usepackage{fancyhdr}

\makeindex

\newboolean{BoolAllPublic}
\setboolean{BoolAllPublic}{false}

\def\overstrike#1#2{{\setbox0\hbox{$#2$}\hbox to \wd0{\hss
    $#1$\hss}\kern-\wd0\box0}}
\def\opm{~\overstrike{\bigcirc}{\pm}~}
\def\omp{~\overstrike{\bigcirc}{\mp}~}

\begin{document}

\title{Few Cycle Optical Pulse Propagation: a detailed calculation}
\author{Paul Kinsler}
\affiliation{
  Department of Physics$^*$, Imperial College,
  Prince Consort Road,
  London SW7 2BW, 
  United Kingdom.
}

\lhead{\includegraphics[height=5mm,angle=0]{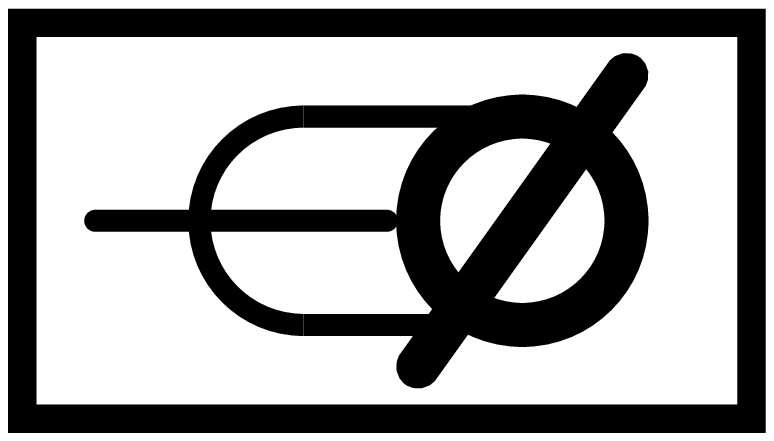}~~(arXiv.org version)}
\chead{~}
\rhead{
\href{mailto:Dr.Paul.Kinsler@physics.org}{Dr.Paul.Kinsler@physics.org}\\
\href{http://www.kinsler.org/physics/}{http://www.kinsler.org/physics/}
}

\date{\today}
\maketitle
\thispagestyle{fancy}

\newcommand{\xxref}[1]{{\bf SeeNote:#1:}}
\newcommand{\xxlabel}[1]{[{\bf SeeRef:{#1}}]}

This document contains my detailed calculation of the Generalised 
Few-cycle Envelope Approximation (GFEA) propagation equation 
reported and used in Phys. Rev. A67, 023813 (2003) 
 \cite{Kinsler-N-2003pra}
and its associated longer version at arXiv.org \cite{Kinsler-N-2002longFCOPO}.
This GFEA propagation equation is intended to be applicable
to optical pulses only a few cycles long, a regime where the standard
Slowly Varying Envelope Approximation (SVEA) fails.

The calculation is intended to be as complete as possible, but is still 
a ``work in progress'', and so may, despite my best efforts, contain 
occaisional mistakes.   It is an edited version of a longer document
from which on-going work has been excised.
Please contact me if you have any comments, corrections
or queries.

[*] I worked at this institution while doing the bulk of this 
calculation.  My main project was with Prof. G.H.C. New on 
few-cycle optical pulses, and I was funded with money from the EPSRC.

{\Large
IMPORTANT NOTE: this calculation is of historical interest only:
the current state of the art is summarized in arXiv:0707.0982.
Please note that the approach taken here is now entirely redundant.
}

~

\noindent
WWW: QOLS Group 
\href{http://www.qols.ph.ic.ac.uk/}{http://www.qols.ph.ic.ac.uk/} \\
WWW: Physics Dept. 
\href{http://www.ph.ic.ac.uk/}{http://www.ph.ic.ac.uk/} \\
WWW: Imperial College
\href{http://www.ic.ac.uk/}{http://www.ic.ac.uk/} \\
Email: Paul Kinsler 
\href{mailto:Dr.Paul.Kinsler@physics.org}{Dr.Paul.Kinsler@physics.org}\\
Email: G.H.C. New
\href{mailto:g.new@ic.ac.uk}{g.new@ic.ac.uk} \\

~
{
\tiny
 \$Id: fewcyc.tex,v 1.3 2007/04/10 11:04:00 physics Exp physics \$
}

\tableofcontents
\newpage

\chead{Few Cycle Pulse Propagation}


\section{Introduction}

This report is a calculation intended to generate an evolution equation 
for an \index{envelope}envelope approximation description of pulse propagation in the
few-cycle regime.  I try to
fully investigate each step and approximation, include all the algebra,
and discuss any subtleties that may arise.  

The calculation parallels and extends that in $\bigstar \bigstar$ T. Brabec,
F. Krausz, {\it ``Nonlinear optical pulse propagation in the single-cycle
regime''} \cite{Brabec-K-1997prl}.  I also (amongst other things) make some 
comments about their slowly-evolving-wave-approximation (SEWA).  

The calculation includes all the steps taken from the
Brabec and Krausz\index{Brabec and Krausz}\cite{Brabec-K-1997prl} starting point of (their eqn.(1))

\begin{eqnarray}
  \left( \partial_z^2 + \nabla_\bot^2 \right) E(\vec{r},t)
- \frac{1}{c^2}
  \partial_t^2
  \int_{-\infty}^t dt' \epsilon(t-t') E(\vec{r},t')
&=&
  \frac{4\pi}{c^2}
  \partial_t^2
  P(\vec{r},t)
,
\label{startingpoint}
\end{eqnarray}

through to their basic \index{nonlinear}nonlinear \index{envelope}envelope equation (NEE) (their eqn.(6))

\begin{eqnarray}
\partial_\xi A
&=&
-\frac{\alpha_0}{2} A
+ \imath \hat{D} A
+ \frac{\imath}{2\beta_0}
  \left( 1 + \frac{\imath}{\omega_0}\partial_\tau \right)^{-1}
  \nabla_\bot^2 
  A
+ \imath \frac{2 \pi \beta_0}{n_0^2}
  \left( 1 + \frac{\imath}{\omega_0}\partial_\tau \right)
  B
.
\label{BrabecKrauszNEE}
\end{eqnarray}

This report also derives the NEE equation found by $\bigstar$ M.A. Porras,
{\it ``Propagation of single-cycle pulsed light beams in dispersive media''}
\cite{Porras-1999pra} which extends the Brabec and Krausz\index{Brabec and
Krausz}\cite{Brabec-K-1997prl} theory so as to treat diffraction and self-focussing better, but which
neglects the \index{nonlinear}nonlinearity, giving their SEEA equation:

\begin{eqnarray}
\partial_\xi
A
&=&
-\frac{\alpha_0}{2} A
+ \imath \hat{D} A
+ \frac{\imath}{2\beta_0}
  \left( 1 + \frac{\imath \beta_1}{\beta_0}\partial_\tau \right)^{-1}
  \nabla_\bot^2 
  A
.
\label{PorrasNEE}
\end{eqnarray}

There seem to be hints of the Brabec and 
Krausz\index{Brabec and Krausz}\cite{Brabec-K-1997prl} result in the early 
paper by $\bigstar$
J.A. \index{Fleck}Fleck, {\it ``Ultra-short pulse
generation by Q-sqitched lasers''} \cite{Fleck-1970prb}.  Fleck\index{Fleck}\cite{Fleck-1970prb} has an \index{envelope}envelope eqn.(2.3a,b) of (not using eqn.(2.2))

\begin{eqnarray}
\frac{\eta}{c}\partial_t \mathcal{E}^+
+
\partial_z \mathcal{E}^+
&=&
 - \left( 1 + \imath \delta \right) P' + {\rm c.c.}
,\\
\frac{\eta}{c}\partial_t \mathcal{E}^-
-
\partial_z \mathcal{E}^-
+ 2 \imath k
&=&
 - \left( 1 + \imath \delta \right) P' + {\rm c.c.}
,
\label{FleckEE}
\end{eqnarray}

where the $1 + \imath \delta$ results from a $1 + \partial_t/\omega$ term
acting on the polarization, which has a linearly decaying memory with
characteristic time $T_2$ (and $\delta=\left(\omega T_2\right)^{-1}$ ); 
dispersion and diffraction are not included.  It is seems plausible that
we can get from eqn (\ref{BrabecKrauszNEE}) to (\ref{FleckEE})
-- we scale the variables (the $\partial_t$ will disappear with a co-moving
frame), neglect diffraction, neglect dispersion, and introduce the same
polarization model. Etc.  This will be clearer as we follow the full
calculation in the next sections.  

An early review of attempts to describe ultra-short pulse
propagation was given by Lamb \cite{Lamb-1971rmp}.

~





\section{Envelopes and Carriers}\label{s-envcarriers}

The substitution used in the ``\index{envelope}envelope approximation'' is
just a splitting of a general waveform (of e.g. the electric field amplitude)
into two parts, an ``\index{envelope}envelope'', and a
``\index{carrier}carrier''; where the \index{carrier}carrier part is intended
to carry almost all of the oscilliatory part of the waveform.  It is usual to
describe a plane-polarised wave, with the $B$ field perpendicular to, in phase
with, and proportional to (in amplitude) the $E$ field.  It is also possible to
do such a calculation with circularly polarized field variables
\cite{Kanetsyan-2002iqec} (using $E_c = E_x + \imath E_y$ rather than just
$E_x$).

For a forwardly propagating \index{carrier}carrier $\exp \left[ \imath \left(
\beta_0 z - \omega_0 t + \psi_0 \right) \right]$ and an
\index{envelope}envelope $A(\vec{r}_\bot,z,t)$, the substitution is

\begin{eqnarray}
  E(\vec{r},t) 
&=& 
  A(\vec{r}_\bot,z,t) 
  e^{\imath \left( \beta_0 z - \omega_0 t + \psi_0\right) } 
 +
  A^*(\vec{r}_\bot,z,t) 
  e^{-\imath \left( \beta_0 z - \omega_0 t + \psi_0\right) } 
\\
&=&
  A(\vec{r}_\bot,z,t) 
  e^{\imath\Xi^-}
 +
  A^*(\vec{r}_\bot,z,t) 
  e^{-\imath \Xi^-}
,
\label{carrierenvelope}
\\
\Xi^\mp &=& \left( \beta_0 z \mp \omega_0 t + \psi_0\right)
.
\label{e-carrier}
\end{eqnarray}


Note that the \index{carrier}carrier is forward propagating because of the
chosen signs on the wavevector ($\beta_0$) and frequency ($\omega_0$) parts:
both terms in eqn.(\ref{carrierenvelope}) are forward propagating (c.f.  a
wave described by e.g. $f(x-vt)$).  Also note that the Poynting vector for the
\index{carrier}carrier field also has a direction, given by $E \times B$ --
and here $B$ is determined by $E$ (c.f. 
Fleck\index{Fleck}\cite{Fleck-1970prb}'s approach).  Using
a forward \index{carrier}carrier means that any backward propagating
components that happen to be in the $E$ waveform need to be contained in the
\index{envelope}envelope -- unless  extra backward \index{carrier}carrier
terms are added to the substitution above.  However, since in many useful
situations an initially forwardly propagating wave does not develope a
significant backward propagating component, we can use approximations to work
in a regime where backward contributions are negligible, rather than
complicate our representation of the $E$ waveform: a forward+backward
\index{carrier}carrier equation might look like this:

\begin{eqnarray}
  E(\vec{r},t) 
&=& 
  A(\vec{r}_\bot,z,t) 
  e^{\imath \left( \beta_0 z - \omega_0 t + \psi_{+}\right) } 
 +
  A^*(\vec{r}_\bot,z,t) 
  e^{-\imath \left( \beta_0 z - \omega_0 t + \psi_{+}\right) } 
\nonumber
\\
& & 
 +
  B(\vec{r}_\bot,z,t) 
  e^{\imath \left( \beta_0 z + \omega_0 t + \psi_{-}\right) } 
 +
  B^*(\vec{r}_\bot,z,t) 
  e^{-\imath \left( \beta_0 z + \omega_0 t + \psi_{-}\right) } 
.
\label{e-doublecarrierenvelope}
\end{eqnarray}

Typically we then try to factor the \index{carrier}carrier part out of our
equations of motion for the waveform, and simplify the
\index{envelope}envelope equation of motion by making approximations based on
assumptions of (e.g.) smoothness of the \index{envelope}envelope, the small
contributions from backward propagating terms, and so on.  This then leaves us
with a (hopefully) manageable equation for just the \index{envelope}envelope
function.   Note that our choice of \index{carrier}carrier frequency is only
constrained by the need to keep the approximations manageable.  

One point with the use of the \index{envelope}envelope function substitution
is that any phase-like properties of the waveform become obscured.  This is
because they can be contained in either (or both) the \index{carrier}carrier
{\em and} the \index{envelope}envelope.  An alternative choice in the phase of
the \index{carrier}carrier (e.g. replacing $\psi_0$ with some $\psi'_0 \neq
\psi_0$) will in fact mean that the \index{envelope}envelope function is
different.  Plotted on a graph, this can seem to have a large effect,
particularly for a few-cycle pulse, where the peak of the waveform moves about
noticeably if the phase of the \index{carrier}carrier is changed in a
\index{envelope}envelope-\index{carrier}carrier pair -- see fig. 1 and 
comments in Brabec and
Krausz\index{Brabec and Krausz}\cite{Brabec-K-1997prl} for some
analysis.

Note however that the \index{envelope}envelope (and \index{carrier}carrier)
are complex -- and a phase shift of $\exp \left[\imath
\left(\psi'_0-\psi_0\right) \right]$ in the \index{carrier}carrier can be
exactly matched by a fixed shift of $\exp \left[-\imath
\left(\psi'_0-\psi_0\right) \right]$ in the \index{envelope}envelope.  This
means there is no calculational or simulation problem associated with shifting
the \index{carrier}carrier phase, as long as the \index{envelope}envelope is
also adjusted.  What matters is that the real $E$ field waveform resulting
from a \index{envelope}envelope-carrier pair is correct according to the
boundary/initial conditions.  Confusion can only occur from the point of view
of an \index{envelope}envelope-only picture of the pulse shape.

\begin{figure}
\includegraphics[width=80mm]{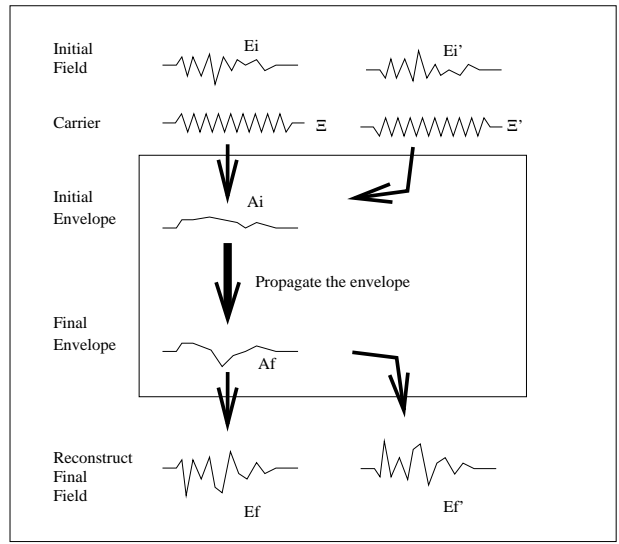}
\caption{
\label{F-envcar}
{\bf F-envcar}:
Diagram showing how a field $E_i$ is decomposed into an 
\index{envelope}envelope $A_i$ and 
a \index{carrier}carrier $\Xi$, the \index{envelope}envelope $A$ is 
propagated to its final state $A_f$, 
then the final field $E_f$ is reconstructed.  It also shows that an alternate
field $E_i'$ can have the same initial \index{envelope}envelope when a 
different \index{carrier}carrier
$\Xi'$ is used, and that the same propagation can be used to extract $E_i$'s
final state $E_f'$.
}
\end{figure}

In the usual case of a pulse that is long compared to its natural
\index{carrier}carrier frequency, the \index{envelope}envelope should be
smooth and so the relationship between the underlying \index{carrier}carrier
{\em phase} and the electric field $E()$ is often ignored, or, if not ignored,
then can be regarded as specified by the many oscillations in $E$ at the
\index{carrier}carrier frequency.  This tend to lead one towards an
``\index{envelope}envelope only'' view of the pulse, which can later cause
confusion for few-cycle pulses where a different choice of
\index{carrier}carrier phase leads to a different looking
\index{envelope}envelope on a $E$ vs $z$ (or $t$) graph, where only the real
part is plotted.  For some kind of complex plot, we would clearly see that
these different looking \index{envelope}envelopes are complex rotations of
each other, and their ``differences'' are merely an artifact (but see the 
following subsection).

\subsection{A Phase Function}

Some authors (e.g. \cite{Lamb-1971rmp,Xiao-WX-2002pra}) use a separate 
phase function $\psi(\vec{r},t)$ in order to ensure their
envelope function remains real.  In this case the definition in
eqn.(\ref{e-carrier}) looks like

\begin{eqnarray}
  E(\vec{r},t) 
&=& 
  A_R(\vec{r}_\bot,z,t) 
  e^{\imath \left( \beta_0 z - \omega_0 t + \psi(\vec{r},t)\right) } 
 +
  A_R(\vec{r}_\bot,z,t) 
  e^{-\imath \left( \beta_0 z - \omega_0 t + \psi(\vec{r},t)\right) } 
.
\label{e-carrier-realA}
\end{eqnarray}

Whilst this might look like a good idea, it complicates any kind of 
propagation equation for the pulse envelope that we might derive: it would
now contain additional derivative terms (of the phase function) and 
also we would need a propagation equation for the phase function itself.
Further, this phase function is ambiguous (recall $\psi+\theta$ is the same 
angle as $\theta$), would be undefined (or any value) when $A_R$
is zero, and numerically difficult to handle when $A_R$ is small.

I do not use this sort of phase function in this document.


\subsection{A Single Envelope Represents a Set of Pulses }

For a given set of initial conditions (usually just the electric field profile
of the input pulse(s), we might pick any value of carrier phase we liked. 
Each different value of carrier phase would result in a different pulse
envelope -- but each of the resulting  combinations of carrier phase and
envelope will specify the same initial conditions.

So, starting with a fixed field, you can use a variety of
\index{carrier}carrier choices, and end up with a variety of
\index{envelope}envelopes.  After solving for the propagation of the chosen
``initial state'' envelope, we get a ``final state'' envelope. From this we
can reconstruct a unique final state $E$ field.  This parallels the 
left hans side of Fig. \ref{F-envcar}. 

{\em But} note since a given \index{envelope}envelope may be turned back into
an electric field by applying any values for carrier phase, this final state
envelope can be used to generate a range of final state  electric fields. 
Each of these fields corresponds to the initial condition specified by the 
initial state envelope and that same choice of carrier phase. 

So {\em one} envelope simulation provides a range of
$E_i \rightarrow E_f$ solutions, as indicated diagramatically on fig. 
\ref{F-envcar}. 

Note that for polarization terms with their own dynamics (e.g. a two level 
atom, see my report \href{twolevelatom.dvi}{\em Two level atoms and the few 
cycle regime} \cite{Kinsler-TLAFC}), the choice of carrier phase alters not 
only the pulse
envelopes but also the representation of the initial polarization state.


\subsection{The so-called ``carrier phase'', i.e. the pulse phase}

Many authors publishing on short pulses refer to ``the carrier phase''. By
this they seem to mean something derived by comparing an inferred field
envelope to the peaks and troughs of the actual (oscillating) electric field. 
For example, they might take the distance between the peak of the envelope as
their zero (i.e. the point of reference), and the peak of the nearest electric
field oscillation as giving  the field phase (see fig \ref{fig-CP}).  A good
example of this is the paper by Chelkowski and Bandrauk
\cite{Chelkowski-B-2002pra}.  To work, this ``maximum amplitude'' procedure
assumes a number of things:

\begin{figure}
\includegraphics[width=60mm]{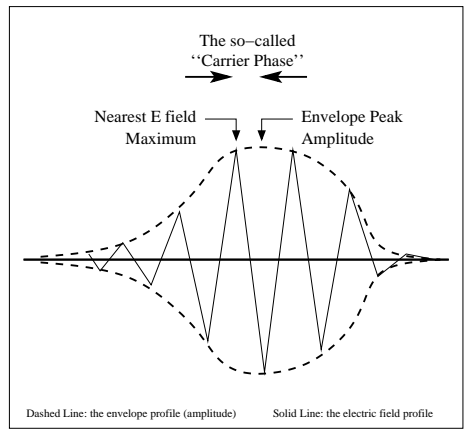}
\caption{Diagram showing traditional ``carrier phase''
estimation method.  The electric field is plotted a sawtooth 
because it was quick to draw -- a modulated sinusoid would be a more 
realistic profile.
\label{fig-CP}
}
\end{figure}

\begin{description}

\item[(a1) The envelope has a single peak:] Of course this is not always true. 
Further, even if it does happen in some particular case to be true, that peak
is not always well localised -- such as in flat-topped pulses.

\item[(a2) The pulse does not have a complex phase structure:] Of course it is
straightforward to generate real valued envelopes from an oscillating electric
field profile, as the electric field is real valued.  However, if (e.g.)  the
pulse is chirped, the electric field oscillations will no longer have a fixed
frequency, so it is not possible to use and envelope-peak to field-peak
distance to guess a phase without additional assumptions.

\end{description}

I believe it is unhelpful to talk of ``carrier phase'' in the contexts the
term is usually applied.  From a mathematical point of view, there are two
phases: the carrier phase, and the envelope phase; further, the envelope phase
may well have a complicated structure that obscures or overrides the role of
the carrier phase.  The carrier phase should be {\em fixed} according to some
spatio-temporal reference point, then further discussions along the lines of
``the phase of the pulse'' should refer to the {\em envelope phase}, and so
the phase structure of the pulse will not so easily be ignored or sidestepped.
However, in the event that a single phase parameter is appropriate or
desirable, a clearly defined method  should be used to extract 
it\ifthenelse{\boolean{BoolAllPublic}{(see following)}}{}, 
avoiding unreliable and subjective
arguments about which value of the phase profile of the envelope is ``the
phase''.




\section{Extending Brabec and Krausz: the post-transform envelope}

Brabec and Krausz\index{Brabec and Krausz}\cite{Brabec-K-1997prl} consider the case of small transverse inhomogeneities of the
polarization, and so start with the three dimensional wave equation

\begin{eqnarray}
  \left( \partial_z^2 + \nabla_\bot^2 \right) E(\vec{r},t)
- \frac{1}{c^2}
  \partial_t^2
  \int_{-\infty}^t dt' \epsilon(t-t') E(\vec{r},t')
&=&
  \frac{4\pi}{c^2}
  \left(
    \partial_t
   \opm
    \frac{gc}{n}
    \partial_z
  \right)
  \partial_t
  P_{nl}(\vec{r},t)
.
\label{wave3d}
\end{eqnarray}

Here (as in Brabec and Krausz\index{Brabec and Krausz}\cite{Brabec-K-1997prl}) $\nabla_\bot^2$ is the transverse Laplace operator,
$\partial_\alpha$ is used as a shorthand notation for $\partial/\partial
\alpha$, $\epsilon(t) = (2\pi)^{-1} \int_{-\infty}^\infty d\omega
\tilde{\epsilon}(\omega) e^{\imath \omega t}$, $\tilde{\epsilon}(\omega) = 1 +
4\pi \chi(\omega)$, and $\chi(\omega)$ is the linear electric susceptibility. 
The electric field $E$ propagates along the $z$ direction. Both $E$ and the
\index{nonlinear}nonlinear polarization $P_{nl}$ are polarized parallel to the $x$ axis.

The $\opm \frac{gc}{n} \partial_z$ is a new term include to allow
the calculation to apply to the $G^\pm$ Fleck\index{Fleck}\cite{Fleck-1970prb} variables.  For 
normal cases set $g=0$ and forget it.  Here I use an alternative ``$\pm$'' 
sign:  $\opm$ -- this is 
to distinguish this variable sign from a later (and independent)
variable sign $\pm$ cause by the choice of \index{carrier}carrier direction.

\subsection{The linear electric susceptibility}

Now we need to treat the effects of the linear electric susceptibility.  We
start by fourier transforming the equation using $\exp\left(-\imath \omega t
\right)$; but neglecting to keep track of the normalisation, since this will
take care of itself when we transform back.  Using the correspondance
$\partial_t \leftrightarrow - \imath \omega$, the first LH term is simple, and
transforms to:

\begin{eqnarray}
  \left( \partial_z^2 + \nabla_\bot^2 \right)
    \tilde{E}(\vec{r}_\bot,z,\omega)
.
\end{eqnarray}
 
The  RH term is also simple, and transforms to:

\begin{eqnarray}
  \frac{4\pi}{c^2}
  \left( 
    -\imath \omega 
   \opm
    \frac{gc}{n} \partial_z
  \right)
  \left( -\imath \omega \right)
    \tilde{P}_{nl}(\vec{r}_\bot,z,\omega) 
.
\end{eqnarray}

The second LH term (with the $dt'$ integral) is more complicated:

\begin{eqnarray}
&~& 
 -\int_{-\infty}^{+\infty} dt ~
    e^{-\imath \omega t}
      \frac{1}{c^2}
      \partial_t^2
      \int_{-\infty}^t dt' ~
        \epsilon(t-t') 
        E(\vec{r}_\bot,z,t') 
\\
&=&
 -
  \frac{\left( -\imath \omega \right)^2}{c^2}
  \int_{-\infty}^{+\infty} dt ~
    e^{-\imath \omega t}
      \int_{-\infty}^t dt' ~ 
        \epsilon(t-t')
        E(\vec{r}_\bot,z,t') 
.
\end{eqnarray}

If the  upper limit of the $dt'$ integral was $t$ and not $\infty$, the $dt'$
integral part would be the normal convolution integral; hence we could convert
it into the product of the fourier transforms of its constituents.  This could
be justified by saying $\epsilon$ must be causal, and so is $0$ for any $t' >
t$, hence the limits of the integral can be extended to $+\infty$.  

\ifthenelse{\boolean{BoolAllPublic}}
{
{\bf Note:}
 {\it
Something seems to be missed in the reverse transform: it does 
not know about the $t$ cut-off, and hence may (will?) not provide a true
reconstruction of $\epsilon(t)$ from $\tilde{\epsilon}(\omega)$.  This
step has therefore introduced a partially acausal $\epsilon(t)$.}  
Presumably this is a known problem ... find a suitable reference?  Maybe 
some kind of material response time could solve the problem?
}

Extending the upper limit of the $dt'$ integral to $\infty$ gives

\begin{eqnarray}
&~&
 +
  \frac{\omega^2}{c^2}
  \int_{-\infty}^{+\infty} dt ~
    e^{-\imath \omega t}
      \int_{-\infty}^{+\infty} dt' ~ 
        \epsilon(t-t')
        E(\vec{r}_\bot,z,t') 
\\
&=&
 +
      \frac{\omega^2}{c^2}
  \left\{
    \int_{-\infty}^{+\infty} dt ~
      e^{-\imath \omega t}
          \epsilon(t)
  \right\}
  \left\{
    \int_{-\infty}^{+\infty} dt ~
      e^{-\imath \omega t}
        E(\vec{r}_\bot,z,t) 
  \right\}
\\
&=&
 +
      \frac{\omega^2}{c^2}
        \tilde{\epsilon}(\omega)
        \tilde{E}(\vec{r}_\bot,z,\omega) 
.
\end{eqnarray}

The resulting transformed version of eqn. (\ref{wave3d}) is

\begin{eqnarray}
  \left( \partial_z^2 + \nabla_\bot^2 \right)
    \tilde{E}(\vec{r}_\bot,z,\omega)
 +
      \frac{\omega^2}{c^2}
        \tilde{\epsilon}(\omega)
        \tilde{E}(\vec{r}_\bot,z,\omega) 
&=&
  \frac{4\pi}{c^2}
  \left( 
    -\imath \omega 
   \opm
    \frac{gc}{n} \partial_z
  \right)
  \left( -\imath \omega \right)
    \tilde{P}_{nl}(\vec{r}_\bot,z,\omega) 
.
\end{eqnarray}

Now I might want to expand $\tilde{\epsilon}(\omega)$ in powers of $\omega$,
but to make things easier I'll replace it with the $k^2(\omega)$ and expand $k$
about $\omega_\epsilon$ instead.  Using $\tilde{\epsilon}(\omega) = c^2
k(\omega)^2 / \omega^2$ (as do Brabec and Krausz\index{Brabec and Krausz}\cite{Brabec-K-1997prl}) and then 

\begin{eqnarray}
 k(\omega)&=&\sum_{n=0}^\infty 
             \frac{\gamma_n \left( \omega - \omega_\epsilon \right)^n}
                  {n!}
; ~~~
  \gamma_n  = \left. \partial_\omega^n k(\omega) \right|_{\omega_\epsilon}
            = \beta_n + \imath \alpha_n
; ~~~ 
  \beta_n, \alpha_n \in \mathbb{R}
; ~~~
  \alpha_n  = \mathbb{I}\mathrm{m}(\gamma_n)
,
  \beta_n   = \mathbb{R}\mathrm{e}(\gamma_n)
.
\end{eqnarray}

Note that Brabec and Krausz\index{Brabec and Krausz}\cite{Brabec-K-1997prl}
have $\alpha_n/2$ in their equations where my definitions will give
$\alpha_n$.   This is because the definition Brabec and Krausz\index{Brabec
and Krausz}\cite{Brabec-K-1997prl} give for $\alpha_n$ (below their eqn.
(BK3)) is not the one they actually use.  The one they use is consistent with
$\alpha$ corresponding to the decay in the intensity, not the field; and in
fact Porras\index{Porras}\cite{Porras-1999pra} alters his definition of $\alpha_n$
from that stated by Brabec and Krausz\index{Brabec and
Krausz}\cite{Brabec-K-1997prl} (and me) in order to have terms like
$\alpha_n/2$ appear.  Porras\index{Porras}\cite{Porras-1999pra}, despite his
different definition, is consistent with my calculations.  The $\alpha_n$ that
Brabec and Krausz\index{Brabec and Krausz}\cite{Brabec-K-1997prl} use is the
same as that defined by Porras\index{Porras}\cite{Porras-1999pra}; and both
Brabec and Krausz\index{Brabec and Krausz}\cite{Brabec-K-1997prl} and
Porras\index{Porras}\cite{Porras-1999pra} use $\omega_\epsilon=\omega_0$.

Using this expansion, the equation becomes

\begin{eqnarray}
  \left( \partial_z^2 + \nabla_\bot^2 \right)
    \tilde{E}(\vec{r}_\bot,z,\omega)
 +
      \frac{\omega^2}{c^2}
      \frac{c^2}{\omega^2}
        k(\omega)^2
        \tilde{E}(\vec{r}_\bot,z,\omega) 
&=&
  \frac{4\pi}{c^2}
  \left( 
    -\imath \omega 
   \opm
    \frac{gc}{n} \partial_z
  \right)
  \left( -\imath \omega \right)
    \tilde{P}(\vec{r}_\bot,z,\omega)
\\ 
  \left( \partial_z^2 + \nabla_\bot^2 \right)
    \tilde{E}(\vec{r}_\bot,z,\omega)
 +
     \left[
       \sum_{n=0}^\infty 
          \frac{\gamma_n \left( \omega - \omega_\epsilon \right)^n}{n!}
     \right]^2
        \tilde{E}(\vec{r}_\bot,z,\omega) 
&=&
  \frac{4\pi}{c^2}
  \left( 
    -\imath \omega 
   \opm
    \frac{gc}{n} \partial_z
  \right)
  \left( -\imath \omega \right)
    \tilde{P}_{nl}(\vec{r}_\bot,z,\omega) 
\label{eqn-frequencywaveeqn}
.
\end{eqnarray}

This can then be transformed back into the time domain (NB: $\partial_t
\leftrightarrow - \imath \omega$, $\imath \partial_t \leftrightarrow \omega$;
$ (\omega-\omega_\epsilon) \rightarrow (\imath \partial_t + \imath^2
\omega_\epsilon) = \imath ( \partial_t + \imath \omega_\epsilon )$)

\begin{eqnarray}
  \left( \partial_z^2 + \nabla_\bot^2 \right)
    E(\vec{r}_\bot,z,t)
 +
     \left[
       \sum_{n=0}^\infty 
       \frac{\imath^n \gamma_n 
                      \left( \partial_t +\imath \omega_\epsilon 
                      \right)^n
            }
            {n!}
     \right]^2
        E(\vec{r}_\bot,z,t) 
&=&
  \frac{4\pi}{c^2}
  \left( 
    \partial_t
   \opm
    \frac{gc}{n} \partial_z
  \right)
  \partial_t
    P_{nl}(\vec{r}_\bot,z,t) 
\label{eqn-timewaveeqn}
.
\end{eqnarray}

\subsection{The envelope and carrier}

Now I split the field up into an \index{envelope}envelope part and a forwardly
propagating \index{carrier}carrier-wave part using the substitution

\begin{eqnarray}
  E(\vec{r},t) 
&=& 
  A(\vec{r}_\bot,z,t) 
  e^{\imath \left( \beta_0 z \mp \omega_0 t + \psi_0\right) } 
 +
  A^*(\vec{r}_\bot,z,t) 
  e^{-\imath \left( \beta_0 z \mp \omega_0 t + \psi_0\right) } 
\\
&=& 
  A(\vec{r}_\bot,z,t) 
  e^{\imath \Xi^\mp}
 +
  A^*(\vec{r}_\bot,z,t) 
  e^{-\imath \Xi^\mp}
\label{e-envelopecarrier}
,
\end{eqnarray}

and similarly for $P_{nl}(\vec{r},t) = B(\vec{r}_\bot,z,t ; A) e^{\imath
\Xi^\mp} + B^*(\vec{r}_\bot,z,t ; A) e^{-\imath \Xi^\mp}$.  The symbol
$\Xi^\mp$ is introduced purely as a convenient shorthand notation for the
terms in the exponential; and the minus sign (i.e. $\Xi^-$) refers to a
forwardly propagating \index{carrier}carrier, and plus sign (i.e. $\Xi^+$) a
backwardly propagating \index{carrier}carrier.  With these
\index{envelope}envelope-\index{carrier}carrier substitutions, the equation of
motion becomes

\begin{eqnarray}
& &
  e^{\imath \Xi^\mp}
  \left( \left[\imath \beta_0 + \partial_z \right]^2 + \nabla_\bot^2 \right)
    A(\vec{r}_\bot,z,t)
 +
  e^{-\Xi^\mp}
  \left( \left[-\imath \beta_0 + \partial_z \right]^2 + \nabla_\bot^2 \right)
    A^*(\vec{r}_\bot,z,t)
\nonumber
\\
&+&
  e^{\imath \Xi^\mp}
     \left[
       \sum_{n=0}^\infty 
          \frac{\imath^n \gamma_n}{n!}
          \left(\mp \imath \omega_0 + \partial_t + \imath \omega_\epsilon\right)^n
     \right]^2
        A(\vec{r}_\bot,z,t) 
 +
  e^{-\imath \Xi^\mp}
     \left[
       \sum_{n=0}^\infty 
          \frac{ \imath^n \gamma_n}{ n!}
          \left(\pm \imath \omega_0 + \partial_t - \imath \omega_\epsilon\right)^n
     \right]^2
        A^*(\vec{r}_\bot,z,t) 
\nonumber
\\
& & ~~~ = 
  e^{\imath \Xi^\mp}
  \frac{4\pi}{c^2}
  \left[
    \left(\mp \imath \omega_0 + \partial_t \right)
   \opm
    \frac{gc}{n} 
    \left(\imath \beta_0 + \partial_z \right)    
  \right]
  \left(\mp \imath \omega_0 + \partial_t \right)
    B(\vec{r}_\bot,z,t ; A) 
\nonumber
\\
& & ~~~ 
   +
  e^{-\imath \Xi^\mp}
  \frac{4\pi}{c^2}
  \left[
    \left(\pm \imath \omega_0 + \partial_t \right)
   \opm
    \frac{gc}{n} 
    \left(-\imath \beta_0 + \partial_z \right)    
  \right]
  \left(\pm \imath \omega_0 + \partial_t \right)
    B^*(\vec{r}_\bot,z,t ; A) 
\label{eqn-wave-ec}
,
\\
&& 
\left( \textrm{NB: }
          \imath^n
          \left(
             \mp \imath \omega_0 + \partial_t + \imath \omega_\epsilon
          \right)^n
\rightarrow
          \imath^n
          \left(
            \imath \omega_\epsilon \mp \imath \omega_0 + \partial_t
          \right)^n
\rightarrow
          \imath^{2n}
          \left(\omega_\epsilon \mp \omega_0 - \imath \partial_t\right)^n
\rightarrow
          \left(- \omega_0 \right)^n 
          \left( \Upsilon^\mp - \imath \partial_t / \omega_0 \right)^n
\right)
\nonumber
\\
\Longrightarrow
& &
  e^{\imath \Xi^\mp}
  \left( \left[\imath \beta_0 + \partial_z \right]^2 + \nabla_\bot^2 \right)
    A(\vec{r}_\bot,z,t)
 +
  e^{-\imath \Xi^\mp}
  \left( \left[-\imath \beta_0 + \partial_z \right]^2 + \nabla_\bot^2 \right)
    A^*(\vec{r}_\bot,z,t)
\nonumber
\\
&+&
  e^{\imath \Xi^\mp}
     \left[
       \sum_{n=0}^\infty 
          \frac{ \gamma_n \left(-\omega_0\right)^n}{ n!}
          \left( \Upsilon^\mp - \frac{ \imath}{\omega_0} \partial_t \right)^n
     \right]^2
        A(\vec{r}_\bot,z,t) 
 +
  e^{-\Xi^\mp}
     \left[
       \sum_{n=0}^\infty 
          \frac{ \gamma_n  \left(-\omega_0\right)^n}{ n!}
          \left(- \Upsilon^\mp - \frac{ \imath}{\omega_0} \partial_t \right)^n
     \right]^2
        A^*(\vec{r}_\bot,z,t) 
\nonumber
\\
& & ~~~ =
  e^{\imath \Xi^\mp}
  \frac{4\pi}{c^2}
  \left[
    \left(\mp \imath \omega_0 + \partial_t \right)
   \opm
    \frac{gc}{n} 
    \left(\imath \beta_0 + \partial_z \right)    
  \right]
  \left(\mp \imath \omega_0 + \partial_t \right)^2
    B(\vec{r}_\bot,z,t ; A) 
\nonumber
\\
& & ~~~ 
   +
  e^{-\imath \Xi^\mp}
  \frac{4\pi}{c^2}
  \left[
    \left(\pm \imath \omega_0 + \partial_t \right)
   \opm
    \frac{gc}{n} 
    \left(-\imath \beta_0 + \partial_z \right)    
  \right]
  \left(\pm \imath \omega_0 + \partial_t \right)
    B^*(\vec{r}_\bot,z,t ; A) 
,
\label{foreback}
\end{eqnarray}

Here $\Upsilon^\mp = \left( \omega_\epsilon \mp \omega_0 \right) / \omega_0$,
which is a quantity which usually would be set to zero -- but retaining it
allows me to expand the dispersion around a frequency other than $\omega_0$. 
Note the usage of $\Upsilon^\mp$ and $\omega_\epsilon$ is clumsy, because we
need to alter its sign under complex conjugation; carrier direction reversal
is taken care of with the $\pm$ notation.

I now split eqn (\ref{foreback}) into two separate equations, the first ``$A$
equation'' containing the terms like $e^{\imath \Xi^\mp}$, and the second
``$A^*$ equation'' containing the terms like $e^{-\imath \Xi^\mp}$.  These two
equations are simply the complex conjugates of one another, and so writing
down only the first one is sufficient:

\begin{eqnarray}
&&
  e^{\imath \Xi^\mp}
  \left( \left[\imath \beta_0 + \partial_z \right]^2 + \nabla_\bot^2 \right)
    A(\vec{r}_\bot,z,t)
 +
  e^{\imath \Xi^\mp}
     \left[
       \sum_{n=0}^\infty 
          \frac{ \gamma_n \left(-\omega_0\right)^n}{ n!}
          \left( \Upsilon^\mp - \frac{ \imath}{\omega_0} \partial_t \right)^n
     \right]^2
        A(\vec{r}_\bot,z,t) 
\\ 
&& ~~~~ = 
  e^{\imath \Xi^\mp}
  \frac{4\pi}{c^2}
  \left[
    \left(\mp \imath \omega_0 + \partial_t \right)
   \opm
    \frac{gc}{n} 
    \left(\imath \beta_0 + \partial_z \right)    
  \right]
  \left(\mp \imath \omega_0 + \partial_t \right)
    B(\vec{r}_\bot,z,t ; A) 
.
\end{eqnarray}

This is simplified with number of minor steps: 
dividing by the $e^{\imath \Xi^\mp}$ factors (which are conveniently never
zero), 
extracting factors of $\mp \omega_0$ from the RHS 
(  $\Rightarrow (\mp \imath \omega_0 + \partial_t) 
= (\mp \imath \omega_0 \mp \imath \omega_0 \partial_t / (\mp \imath \omega_0)) 
= \mp \imath \omega_0 ( 1 + \partial_t / (\mp \imath \omega_0) 
= \mp \imath \omega_0 ( 1 \mp \partial_t / (\imath \omega_0) 
= \mp \imath \omega_0 ( 1 \pm \imath \partial_t / \omega_0$),
 then preparing to use $1 = c \beta_0 / n \omega_0$, leaves

\begin{eqnarray}
&&
  \left( \left[\imath \beta_0 + \partial_z \right]^2 + \nabla_\bot^2 \right)
    A(\vec{r}_\bot,z,t)
 +
     \left[
       \sum_{n=0}^\infty 
          \frac{\gamma_n \left(-\omega_0\right)^n}{ n!}
          \left( \Upsilon^\mp - \frac{\imath}{\omega_0} \partial_t \right)^n
     \right]^2
        A(\vec{r}_\bot,z,t) 
\\ 
&& ~~~~ = 
 -
  \frac{4 \pi \omega_0^2}{c^2}
  \left[
    \left( 1 \pm \frac{\imath}{\omega_0}\partial_t \right)
   \omp
    \frac{g c \beta_0}{n \omega_0} 
    \left(1 - \frac{\imath}{\beta_0} \partial_z \right)    
  \right]
  \left(1 \pm \frac{\imath}{\omega_0}\partial_t \right)
    B(\vec{r}_\bot,z,t ; A) 
.
\label{nearlyBKeqn2}
\end{eqnarray}

This appears to differ slightly from Bracbec-Krausz eqn.(2) in that it has
the opposite sign on the RHS -- however, agreement is recovered later in
eqn.(\ref{exact-BKP}).

If $A_0$ is a solution of the $A$ equation, then its conjugate $A_0^*$ is a
solution of the conjugate $A^*$ equation.  This means that solving one solves the other, and a total
waveform can then be easily reconstructed using eqn (\ref{e-envelopecarrier}). 
There are {\em no} approximations made in doing this, but it may be that there
are complicated (or subtle) cases where solutions of the full equation are not
expressible in terms of solutions of the two separate ones.

Note that for the forward propagating \index{carrier}carrier ($\Xi^-$, the
upper sign choice in $\Xi^\mp$), there are no explicitly backward propagating
terms, despite the fact that we have not excluded them in any way -- this is
because they do not arise spontaneously, but  need to be created.  Such an
effect could occur in the case of multi-field systems or exotic polarization
behaviour, where there may well be spatially oscillating terms from the
\index{nonlinear}nonlinear polarization term $B$ ( e.g.  $\exp \left( \pm
\imath k_B z \right)$) which could force a (possibly backward propagating)
oscillation onto $A$.  This would very likely violate some approximation we
will want to make later, e.g.  a ``smooth'' or slowly varying
\index{envelope}envelope function $A$.  We can only neglect backward
propagating components if (a) there were none to start with, and (b) by
verifying (or assuming) that the \index{nonlinear}nonlinear polarization has
convenient properties -- although we could extend eqn
(\ref{e-envelopecarrier}) to include backward \index{carrier}carrier terms as
already discussed (see section \ref{s-envcarriers}).

At the equivalent point to eqn.(\ref{nearlyBKeqn2}) in the Bracbec-Krausz
paper, they already claim to have neglected backward propagating waves (after
their eqn (1)): ``the neglect of backward propagating waves is consistent with
the approximations that will be made in the following derivation of the
\index{envelope}envelope equation and will be commented on later'' -- their
comment being that ``excessive'' change in the \index{envelope}envelope can
lead to backwardly propagating components to the \index{envelope}envelope (see
e.g. Shen \cite{Shen-PNLO}).  I, however, leave any approximations relating to
the neglect of backward terms to later on -- it is still the case that (in
principle) that the \index{envelope}envelope function might contain backwardly
propagating components.

This differs from Fleck\index{Fleck}\cite{Fleck-1970prb} in that his $E^\pm$
and $\mathcal{E}^\pm$ ($G^+$ and $\mathcal{G}^+$ in my notation) are
constructed as {\it explicitly} forward and backward propagating, and hence
should not really be compared directly to my $E$ or $A$.  My $E$ is in fact
Fleck\index{Fleck}\cite{Fleck-1970prb}'s $E^+ + E^-$ ($G^+ + G^-$), and
somewhere in the approximations used to get to my starting point
eqn.(\ref{wave3d}) the magnetic field parts (retained by
Fleck\index{Fleck}\cite{Fleck-1970prb}) have been assumed to be solely
dependent on (deriveable from) the electric field.  See my derivation {\it ``A
second-order wave equation using Fleck field
variables''}\cite{Kinsler-RN-2005pra,Kinsler-FLECK} for more information.

\subsection{Scaled co-moving variables}

I now change into a scaled co-moving reference frame, but one slightly
different to that of Bracbec-Krausz -- the difference being that I scale space
by $\beta_0$ and the time by $\omega_0$ as well as shifting the origin. 
Bracbec-Krausz and Porros use $\xi=z$ and $\tau=t \mp \beta_1 z$, but 
I instead put

\begin{eqnarray}
\xi&=&\beta'_0 z 
,
\\
\tau&=&\omega_0 \left(t \mp \beta'_1 z\right)
\label{coords-coscaled}
.
\end{eqnarray}

Here I use $\beta'_0$ and $\beta'_1$ rather than just $\beta_0$ and $\beta_1$
because it may not always be convenient to use the natural scaled co-moving
reference frame; which might well be the case for solving multi-mode problems.
For an $A^*$ equation, the signs of both $z$ and $t$ should be reversed so
that $\xi_- = - \beta'_0 z$ and $\tau_- = -\omega_0 \left(t \mp \beta'_1
z\right)$.  Do {\em not} think that this $A^*$ frame (for $\Xi^-$, upper sign)
is backwardly propagating, since although $\xi_- \sim -z$, time has also
reversed $\tau_- \sim -t$. The derivatives for the $A$ equation transform like 

\begin{eqnarray}
  \partial_t \equiv \frac{d}{dt} 
&=& 
  \frac{d\tau}{dt} \frac{d}{d\tau}
 +
  \frac{d\xi}{dt} \frac{d}{d\xi} 
= \omega_0 \frac{d}{d\tau}
  \equiv \omega_0 \partial_\tau
\\
  \partial_z \equiv \frac{d}{dz} 
&=& 
  \frac{d\xi}{dz} \frac{d}{d\xi} 
 +
  \frac{d\tau}{dz} \frac{d}{d\tau}
= \beta_0 \frac{d}{d\xi} \mp \omega_0 \beta'_1 \frac{d}{d\tau} 
  \equiv \beta'_0 \partial_\xi \mp \omega_0 \beta'_1 \partial_\tau
,
\end{eqnarray}

(for $A^*$, $\partial_t = -\omega_0 \partial_{\tau_-}$ and $\partial_z = -
\beta'_0 \partial_{\xi_-} \pm \omega_0 \beta'_1 \partial_{\tau_-}$).  The
scaled co-moving $A$ equation is then (with $q=\beta'_0/\beta_0$).  Note that
now a condition that  $\xi \gg 1$ refers to lengths $\gg 1/\beta_0$ (i.e.
``long''), and $\tau \gg 1$ refers to times $\gg 1/\omega_0$ (i.e. ``slow''). 
Similarly, $\xi \ll 1$ refers to lengths $\ll 1/\beta_0$ (i.e. ``short''), and
$\tau \ll 1$ refers to times $\ll 1/\omega_0$ (i.e. ``fast'').  

\begin{eqnarray}
&& 0
\\&=&
  \left\{ 
    \left(  \imath \beta_0 
          + \beta'_0 \partial_\xi 
        \mp \omega_0 \beta'_1 \partial_\tau
    \right)^2 
  + \nabla_\bot^2 
  +
     \left[
       \sum_{n=0}^\infty 
          \frac{\gamma_n \left( -\omega_0 \right)^n }{n!}
          \left( \Upsilon^\mp - \imath \partial_\tau \right)^n
     \right]^2
  \right\}
        A(\vec{r}_\bot,\xi,\tau) 
\nonumber
\\
&& +
  \frac{4 \pi \omega_0^2}{c^2}
  \left[
    \left( 1 \pm \imath \partial_\tau \right)
   \omp
    \frac{g c \beta_0}{n \omega_0} 
    \left(
        1 
      - 
        \imath q \partial_\xi
      - 
        \imath \sigma' \partial_\tau
    \right)
  \right]
  \left( 
         1 \pm \imath \partial_\tau  
  \right)
    B(\vec{r}_\bot,\xi,\tau ; A) 
~~~~ \textrm{ ... now divide through by $\beta_0^2$ ...}
\\
&=&
  \left\{ 
    -
    \left(  1
            - \imath q \partial_\xi 
           \pm \frac{\imath \omega_0 \beta'_1 }{\beta_0} \partial_\tau
    \right)^2 
  + \frac{1}{\beta_0^2}
    \nabla_\bot^2 
  +
    \left[
       \sum_{n=0}^\infty 
          \frac{\gamma_n \left( -\omega_0 \right)^n }{ \beta_0 n!}
          \left(\Upsilon^\mp  - \imath \partial_\tau \right)^n
     \right]^2
  \right\}
        A(\vec{r}_\bot,\xi,\tau) 
\nonumber
\\
&& +
  \frac{4 \pi \omega_0^2}{c^2 \beta_0^2}
  \left[
    \left( 
        1 
       \pm \imath \partial_\tau 
       \opm \imath g \sigma' \partial_\tau
    \right)
   \omp
    g
    \left(
        1 
      - 
        \imath q \partial_\xi
    \right)
  \right]
  \left( 
         1 \pm \imath \partial_\tau  
  \right)
    B(\vec{r}_\bot,\xi,\tau ; A) 
~~~~ \textrm{ ... now expand, prepare for $\hat{D}'$ ...}
\\
&=&
  \left\{ 
    - 1 
    + 2 
      \imath
      \left( q \partial_\xi 
            \mp \frac{\omega_0 \beta'_1 }{\beta_0} \partial_\tau
      \right)
    + 
      \left( q \partial_\xi 
            \mp \frac{\omega_0 \beta'_1 }{\beta_0} \partial_\tau
      \right)^2 
    + \frac{1}{\beta_0^2}
      \nabla_\bot^2 
    +
    \left[
      \frac{\beta_0}{\beta_0}
      +
      \frac{\imath \alpha_0}{\beta_0}
      -
      \frac{ \omega_0 \beta_1}{\beta_0}
          \left(\Upsilon^\mp  - \imath \partial_\tau \right)
      +
        \frac{ \omega_0}{\beta_0}
      \hat{D}
     \right]^2
  \right\}
        A(\vec{r}_\bot,\xi,\tau) 
\nonumber
\\
&& +
  \frac{4 \pi }{n_0^2}
  \left[
        1 
      \pm  \imath \partial_\tau 
      \opm \imath g \sigma' \partial_\tau
   \omp
    g
    \left(
        1 
      - 
        \imath q \partial_\xi
    \right)
  \right]
  \left( 
         1 \pm \imath \partial_\tau  
  \right)
    B(\vec{r}_\bot,\xi,\tau ; A) 
\\
&=&
  \left\{ 
    - 1 
    + 2 
      \imath
      \left( q \partial_\xi 
            \mp \frac{\omega_0 \beta'_1 }{\beta_0} \partial_\tau
      \right)
    + 
      \left( q \partial_\xi 
            \mp \frac{\omega_0 \beta'_1 }{\beta_0} \partial_\tau
      \right)^2 
    + \frac{1}{\beta_0^2}
      \nabla_\bot^2 
    +
    \left[
      1
      +
      \imath
      \left(
        \imath
          \frac{ \omega_0 \beta_1}{\beta_0}
          \left(\Upsilon^\mp  - \imath \partial_\tau \right)
        +
          \frac{\alpha_0}{\beta_0}
        - 
          \imath
          \hat{D}'
      \right)
     \right]^2
  \right\}
        A(...)  
\nonumber
\\
&& +
  \frac{4 \pi }{n_0^2}
  \left[
        1 
      \pm  \imath \partial_\tau 
      \opm \imath g \sigma' \partial_\tau
   \omp
    g
    \left(
        1 
      - 
        \imath q \partial_\xi
    \right)
  \right]
  \left( 
         1 \pm \imath \partial_\tau  
  \right)
    B(\vec{r}_\bot,\xi,\tau ; A) 
~~~~~~~~~~~~~~  {\rm ... ~and ~now ~divide ~by ~} 2 \imath {\rm ...}
\\
&=&
  \left\{ 
      \left( q \partial_\xi 
            \mp \frac{\omega_0 \beta'_1 }{\beta_0} \partial_\tau
      \right)
    + 
      \frac{1}{2\imath}
      \left( q \partial_\xi 
            \mp \frac{\omega_0 \beta'_1 }{\beta_0} \partial_\tau
      \right)^2 
    + \frac{1}{2 \imath \beta_0^2}
      \nabla_\bot^2 
\right.
\nonumber
\\
& &
\left.
    +
      \frac{2\imath}{2\imath}
      \left[
        \imath
          \frac{ \omega_0 \beta_1}{\beta_0}
          \left( \Upsilon^\mp  - \imath \partial_\tau \right)
        +
          \frac{ \alpha_0}{\beta_0}
        - 
          \imath
          \hat{D}'
      \right]
    +
      \frac{\imath^2}{2\imath}
      \left[
        \imath
          \frac{ \omega_0 \beta_1}{\beta_0}
          \left( \Upsilon^\mp - \imath \partial_\tau \right)
        +
          \frac{\alpha_0}{\beta_0}
        - 
          \imath
          \hat{D}'
      \right]^2
  \right\}
        A(\vec{r}_\bot,\xi,\tau) 
\nonumber
\\
&& +
  \frac{2 \pi }{\imath n_0^2}
  \left[
        1 
      \pm  \imath \partial_\tau 
      \opm \imath g \sigma' \partial_\tau
   \omp
    g
    \left(
        1 
      - 
        \imath q \partial_\xi
    \right)
  \right]
  \left( 
         1 \pm \imath \partial_\tau  
  \right)
    B(\vec{r}_\bot,\xi,\tau ; A) 
\label{forBKcomments}
\\
&=&
  \left\{ 
      \left( q \partial_\xi 
            \mp \sigma' \partial_\tau
      \right)
    + 
      \frac{1}{2\imath}
      \left( q \partial_\xi 
            \mp \sigma' \partial_\tau
      \right)^2 
    + \frac{1}{2 \imath \beta_0^2}
      \nabla_\bot^2 
    +
      \left[
        \imath
          \sigma
          \left( \Upsilon^\mp - \imath \partial_\tau \right)
        +
          \frac{ \alpha_0}{\beta_0}
        - 
          \imath
          \hat{D}'
      \right]
\right.
\nonumber
\\
&&    +
\left.
      \frac{\imath}{2}
      \left[
        \imath
          \sigma
          \left( \Upsilon^\mp - \imath \partial_\tau \right)
        +
          \frac{\alpha_0}{\beta_0}
        - 
          \imath
          \hat{D}'
      \right]^2
  \right\}
        A(\vec{r}_\bot,\xi,\tau) 
\nonumber
\\
&& +
  \frac{2 \pi }{\imath n_0^2}
  \left[
        1 
      \pm  \imath \partial_\tau 
      \opm \imath g \sigma' \partial_\tau
   \omp
    g
    \left(
        1 
      - 
        \imath q \partial_\xi
    \right)
  \right]
  \left( 
         1 \pm \imath \partial_\tau  
  \right)
    B(\vec{r}_\bot,\xi,\tau ; A) 
.
\end{eqnarray}

Here I have introduced the dimensionless $\sigma = \omega_0 \beta_1 / \beta_0
= (\omega_0 / \beta_0) / (1/\beta_1) = v_f / v_g $, $\sigma' = \omega_0
\beta'_1 / \beta_0$, and used the fact that the refractive index at $\omega_0$
is $n_0= c \beta_0 / \omega_0$.  I also define a dispersion term $\hat{D}$ in
a similar way to Brabec and Krausz\index{Brabec and
Krausz}\cite{Brabec-K-1997prl}, but instead use a scaled (dimensionless)
version $\hat{D}'$ in following equations:

\begin{eqnarray}
\hat{D}'
=
  \frac{ \omega_0}{\beta_0}
  \hat{D}
&=& 
  -
  \frac{ \omega_0}{\beta_0}
  \left[
    \imath \alpha_1
    \left( \Upsilon^\mp - \imath \partial_\tau \right)
   +
    \sum_{n=2}^\infty 
          \frac{\gamma_n' \left( -\omega_0 \right)^{n-1} }{n!}
          \left( \Upsilon^\mp - \imath \partial_\tau \right)^n
  \right]
,
\end{eqnarray}

with $\gamma_n' = \gamma_n \textrm{~for~} n \ge 2$; otherwise $\gamma_0'  =
0$; $\gamma_1'  = \imath \alpha_1$: the parameters $\alpha_0$, $\beta_0$,
$\beta_1$ are handled separately from $\hat{D}$ because of their important 
role.

\subsection{Aside: Brabec and Krausz approximation criteria}

Brabec and Krausz\index{Brabec and Krausz}\cite{Brabec-K-1997prl} introduce
some criteria designed to motivate approximations to their equations
(BK5a,b,c).  Since I use the same variable names, but differently scaled, I 
here
write the Brabec and Krausz\cite{Brabec-K-1997prl}
criteria in their form on the LHS, and indicate with an arrow my form on the
RHS.

\begin{eqnarray}
(5a) ~~~~~~~~~~~~~~~~~~
  \left| \partial_\xi A \right| 
 \ll 
  \beta_0 \left| A \right|
&\Longrightarrow&
  \left| \partial_\xi A \right| 
 \ll 
  A,
\\
(5b) ~~~~~~~~~~~~~~~~~~
  \left| \partial_\tau A \right| 
 \ll 
  \omega_0 \left| A \right|
&\Longrightarrow&
  \left| \partial_\tau A \right| 
 \ll 
  A,
\\
(5c) ~~~~~~~~~~~~~~~
  \left| \frac{ \beta_0 - \omega_0 \beta_1 }
              {\beta_0}
  \right|
 \ll 
  1
.
&~&
\end{eqnarray}

Note that $g=0$ here.  The motivation for the first two (5a,b) are obvious
from eqn.(\ref{forBKcomments}), in my scaled co-moving frame -- they allow me
to say certain quantities are small, and hence I could choose to neglect them.
In contrast, it is not clear how the third condition will make my equations
simpler beyond removing a single prefactor (since $\left| \omega_0 \beta_1 /
\beta_0 \right| \approx 1$); but $\left| \omega_0 \beta_1 / \beta_0 \right|
\ll 1$ might seem better still, if perhaps not as physically relevant. 
However, Brabec and Krausz\index{Brabec and Krausz}\cite{Brabec-K-1997prl}
collect their terms together differently, and indeed the situation becomes
clearer after I rearrange the eqations.

\subsection{The Generalised Few-Cycle Envelope equation}

I keep $\left(\mp \Upsilon - \imath \partial_\tau \right)$ terms intact because
for the usual case where $\omega_\epsilon=\omega_0$ is chosen, they will
simplify to $-\imath \partial_\tau$.  Note also I have divided the equation
through by $\beta_0^2$ rather than the single $\beta_0$ of Brabec and Krausz\index{Brabec and Krausz}\cite{Brabec-K-1997prl}. 
Still retaining all terms in the equation, I have 

\begin{eqnarray}
0
&=&
  \left\{ 
      \left( q \partial_\xi 
            \mp \sigma' \partial_\tau
      \right)
    + 
      \frac{1}{2\imath}
      \left( q \partial_\xi 
            \mp \sigma' \partial_\tau
      \right)^2 
    + \frac{1}{2 \imath \beta_0^2}
      \nabla_\bot^2 
    +
      \left[
        \imath
          \sigma
          \left( \Upsilon^\mp - \imath \partial_\tau \right)
        +
        \left(
          \frac{ \alpha_0}{\beta_0}
        - 
          \imath \hat{D}'
        \right)
      \right]
\right.
\nonumber
\\
& &
\left.
    +
    \frac{\imath}{2}
      \left[
        \imath
          \sigma
          \left( \Upsilon^\mp - \imath \partial_\tau \right)
        +
        \left(
          \frac{ \alpha_0}{\beta_0}
        - 
          \imath \hat{D}'
        \right)
      \right]^2
  \right\}
        A(\vec{r}_\bot,\xi,\tau) 
\nonumber
\\
&& +
  \frac{2 \pi }{\imath n_0^2}
  \left[
        1 
      \pm  \imath \partial_\tau 
      \opm \imath g \sigma' \partial_\tau
   \omp
    g
    \left(
        1 
      - 
        \imath q \partial_\xi
    \right)
  \right]
  \left( 
         1 \pm \imath \partial_\tau  
  \right)
    B(\vec{r}_\bot,\xi,\tau ; A) 
\\
(a)&=&
  \left\{ 
      \left( q \partial_\xi 
            \mp \sigma' \partial_\tau
      \right)
    + 
      \frac{1}{2\imath}
      \left( q \partial_\xi 
            \mp \sigma' \partial_\tau
      \right)^2 
      +
        \imath
          \sigma
          \left( \Upsilon^\mp - \imath \partial_\tau \right)
      +
          \left(
            \frac{ \alpha_0}{\beta_0}
          - 
            \imath \hat{D}'
          \right)
\right.
\nonumber
\\
& &
\left.
    + \frac{i}{2}
    \left[
      -   \sigma^2
          \left( \Upsilon^\mp - \imath \partial_\tau \right)^2
      +
          2 \imath
          \sigma
          \left( \Upsilon^\mp - \imath \partial_\tau \right)
          \left(
            \frac{ \alpha_0}{\beta_0}
          - 
            \imath \hat{D}'
          \right)
      + 
          \left(
            \frac{ \alpha_0}{\beta_0}
          - 
            \imath \hat{D}'
          \right)^2
     \right]
    + 
      \frac{1}{2 \imath \beta_0^2}
      \nabla_\bot^2 
  \right\}
        A(\vec{r}_\bot,\xi,\tau) 
\nonumber
\\
&& +
  \frac{2 \pi }{\imath n_0^2}
  \left[
        1 
      \pm  \imath \partial_\tau 
      \opm \imath g \sigma' \partial_\tau
   \omp
    g
    \left(
        1 
      - 
        \imath q \partial_\xi
    \right)
  \right]
  \left( 
         1 \pm \imath \partial_\tau  
  \right)
    B(\vec{r}_\bot,\xi,\tau ; A) 
\\
(a')&=&
  \left\{ 
      \left( q \partial_\xi 
            \mp \sigma' \partial_\tau
      \right)
    + 
      \frac{1}{2\imath}
      \left( q \partial_\xi 
            \mp \sigma' \partial_\tau
      \right)^2 
      +
        \imath
          \sigma
          \left( \Upsilon^\mp - \imath \partial_\tau \right)
      -   \frac{i}{2}
          \sigma^2
          \left( \Upsilon^\mp - \imath \partial_\tau \right)^2
\right.
\nonumber
\\
& &
\left.
      -   
          \sigma
          \left( \Upsilon^\mp - \imath \partial_\tau \right)
          \left(
            \frac{ \alpha_0}{\beta_0}
          - 
            \imath \hat{D}'
          \right)
      +
          \left(
            \frac{ \alpha_0}{\beta_0}
          - 
            \imath \hat{D}'
          \right)
      + 
          \frac{i}{2}
          \left(
            \frac{ \alpha_0}{\beta_0}
          - 
            \imath \hat{D}'
          \right)^2
    + 
      \frac{1}{2 \imath \beta_0^2}
      \nabla_\bot^2 
  \right\}
        A(\vec{r}_\bot,\xi,\tau) 
\nonumber
\\
&& +
  \frac{2 \pi }{\imath n_0^2}
  \left[
        1 
      \pm  \imath \partial_\tau 
      \opm \imath g \sigma' \partial_\tau
   \omp
    g
    \left(
        1 
      - 
        \imath q \partial_\xi
    \right)
  \right]
  \left( 
         1 \pm \imath \partial_\tau  
  \right)
    B(\vec{r}_\bot,\xi,\tau ; A) 
\\
(b)&=&
  \left\{ 
      q \partial_\xi 
    \mp 
      \sigma' \partial_\tau
    + 
      \frac{1}{2 \imath} q^2 \partial_\xi^2
    \mp 
      \frac{1}{2 \imath} 2 q \partial_\xi 
                     \sigma' \partial_\tau
    +
      \frac{1}{2 \imath} \sigma'^2 \partial_\tau^2
    +
    \imath \sigma \Upsilon^\mp
    +
    \sigma \partial_\tau
    -
      \frac{\imath \sigma^2}{2}
      \left( \Upsilon^\mp - \imath \partial_\tau \right)^2
    -
    \sigma
    \left( \Upsilon^\mp - \imath \partial_\tau \right)
    \left(
        \frac{\alpha_0}{\beta_0}
      -
        \imath 
        \hat{D}'
     \right)
\right.
\nonumber
\\
& &
\left.
      +
          \left(
            \frac{ \alpha_0}{\beta_0}
          - 
            \imath \hat{D}'
          \right)
      + 
          \frac{i}{2}
          \left(
            \frac{ \alpha_0}{\beta_0}
          - 
            \imath \hat{D}'
          \right)^2
    + 
      \frac{1}{2 \imath \beta_0^2}
      \nabla_\bot^2 
  \right\}
        A(\vec{r}_\bot,\xi,\tau) 
\nonumber
\\
&& +
  \frac{2 \pi }{\imath n_0^2}
  \left[
        1 
      \pm  \imath \partial_\tau 
      \opm \imath g \sigma' \partial_\tau
   \omp
    g
    \left(
        1 
      - 
        \imath q \partial_\xi
    \right)
  \right]
  \left( 
         1 \pm \imath \partial_\tau  
  \right)
    B(\vec{r}_\bot,\xi,\tau ; A) 
\\
(c)&=&
  \left\{
      \left( 1 \pm \imath \sigma' \partial_\tau \right)
      q \partial_\xi
    +
      \imath \sigma \Upsilon^\mp
    + 
      \frac{1}{2\imath}
      \left( q^2 \partial_\xi^2 
            + \sigma'^2 \partial_\tau^2
      \right)
    +
      \left( \sigma \mp \sigma' \right)
      \partial_\tau
    -
    \frac{\imath \sigma^2}{2}
    \left( \Upsilon^\mp - \imath \partial_\tau \right)^2
    -
    \sigma
    \left( \Upsilon^\mp - \imath \partial_\tau \right)
    \left(
        \frac{\alpha_0}{\beta_0}
      -
        \imath 
        \hat{D}'
     \right)
\right.
\nonumber
\\
& &
\left.
      +
          \left(
            \frac{ \alpha_0}{\beta_0}
          - 
            \imath \hat{D}'
          \right)
      + 
          \frac{i}{2}
          \left(
            \frac{ \alpha_0}{\beta_0}
          - 
            \imath \hat{D}'
          \right)^2
    + 
      \frac{1}{2 \imath \beta_0^2}
      \nabla_\bot^2 
  \right\}
        A(\vec{r}_\bot,\xi,\tau) 
\nonumber
\\
&& +
  \frac{2 \pi }{\imath n_0^2}
  \left[
        1 
      \pm  \imath \partial_\tau 
      \opm \imath g \sigma' \partial_\tau
   \omp
    g
    \left(
        1 
      - 
        \imath q \partial_\xi
    \right)
  \right]
  \left( 
         1 \pm \imath \partial_\tau  
  \right)
    B(\vec{r}_\bot,\xi,\tau ; A) 
\label{continuefromhere}
\\
(d)&=&
  \left\{ 
      \left( 1 \pm \imath \sigma' \partial_\tau \right)
      q \partial_\xi
    +
      \imath \sigma \Upsilon^\mp
    + 
      \frac{1}{2\imath}
      \left( q^2 \partial_\xi^2 
            + \sigma'^2 \partial_\tau^2
      \right)
    +
      \left( \sigma \mp \sigma' \right)
      \partial_\tau
    -
    \frac{\imath \sigma^2}{2}
    \left( \Upsilon^\mp - \imath \partial_\tau \right)^2
    -
    \sigma
    \left( \Upsilon^\mp - \imath \partial_\tau \right)
    \left(
        \frac{\alpha_0}{\beta_0}
      -
        \imath 
        \hat{D}'
     \right)
\right.
\nonumber
\\
& &
\left.
    +
    \frac{\imath}{2}
    -
    \frac{\imath}{2}
    \left[
        1
      +
        \imath
        \left(
          \frac{ \alpha_0}{\beta_0}
        - 
          \imath \hat{D}'
        \right)
     \right]^2
    + \frac{1}{2 \imath \beta_0^2}
      \nabla_\bot^2 
  \right\}
        A(\vec{r}_\bot,\xi,\tau) 
\nonumber
\\
&& +
  \frac{2 \pi }{\imath n_0^2}
  \left[
        1 
      \pm  \imath \partial_\tau 
      \opm \imath g \sigma' \partial_\tau
   \omp
    g
    \left(
        1 
      - 
        \imath q \partial_\xi
    \right)
  \right]
  \left( 
         1 \pm \imath \partial_\tau  
  \right)
    B(\vec{r}_\bot,\xi,\tau ; A) 
\label{eqn-smalldispersion}
.
\end{eqnarray}

This equation (\ref{eqn-smalldispersion}) has a nice 
$1+\imath(\alpha_0/\beta_0- \imath \hat{D}')$ term 
which may be useful in a generic ``small dispersion'' case.

However, we want to reach the more typical Brabec and Krausz\index{Brabec and
Krausz}\cite{Brabec-K-1997prl} (or Porras\index{Porras}\cite{Porras-1999pra}) 
form;
and so instead continue with the algebra from eqn.(\ref{continuefromhere}). 
Like Porras\cite{Porras-1999pra}, I retain the natural choice of
$1 + \imath \sigma \partial_\tau$ multiplier for the $\partial_\xi$ term. 
This differs from the $1 + \imath \partial_\tau$ used by Brabec and
Krausz\cite{Brabec-K-1997prl}, and is why
Porras\cite{Porras-1999pra} claims that the Brabec and
Krausz\cite{Brabec-K-1997prl} SEWA equation has the
space-time focussing ``slightly falsified'', mentioning
Rothenburg\cite{Rothenburg-1992ol} -- it is because Brabec and
Krausz\cite{Brabec-K-1997prl} had already introduced
a $\sigma=1$ approximation, and as such the ``missing'' $\sigma$ is an
approximation, not anything ``slightly falsified''.

Now ruthlessly shift all the $\Upsilon^\mp$, $\mathscr{O}(\partial_\xi^2)$ and
$\mathscr{O}(\partial_\tau^2)$ terms to the RHS...

\begin{eqnarray}
&&
  \left\{
      \left( 1 \pm \imath \sigma' \partial_\tau \right)
      q \partial_\xi
    +  
      \left( 
        \sigma \mp \sigma'
      \right) 
      \partial_\tau
    +  
      \left( 1 + \imath \sigma \partial_\tau \right)
      \left( 
        \frac{ \alpha_0}{\beta_0}
      - \imath  \hat{D}'
      \right) 
  \right\}
    A(\vec{r}_\bot,\xi,\tau) 
  +
    \frac{1}{2 \imath \beta_0^2}
      \nabla_\bot^2 
    A(\vec{r}_\bot,\xi,\tau) 
\nonumber
\\
&& +
  \frac{2 \pi }{\imath n_0^2}
  \left[
        1 
      \pm  \imath \partial_\tau 
      \opm \imath g \sigma' \partial_\tau
   \omp
    g
    \left(
        1 
      - 
        \imath q \partial_\xi
    \right)
  \right]
  \left( 
         1 \pm \imath \partial_\tau  
  \right)
    B(\vec{r}_\bot,\xi,\tau ; A) 
\nonumber
\\
&=&
  -
  \left[
      \imath \sigma \Upsilon^\mp
    -
      \sigma  \Upsilon^\mp 
        \left(
          \frac{ \alpha_0}{\beta_0}
        - 
          \imath \hat{D}'
        \right)
    -
      \frac{\imath \sigma^2}
           {2}
        \left(
          \Upsilon^\mp - \imath \partial_\tau
        \right)^2
    +
      \frac{\imath}{2}
        \left(
          \frac{ \alpha_0}{\beta_0}
        - 
          \imath \hat{D}'
        \right)^2
    +
      \frac{1}{2\imath}
        \left(
          q^2 \partial_\xi^2 
        +
          \sigma'^2 \partial_\tau^2
        \right)^2
  \right]
  A(\vec{r}_\bot,\xi,\tau) 
.
\end{eqnarray}

Presumably Brabec and Krausz\index{Brabec and Krausz}\cite{Brabec-K-1997prl}
used their eqn. (BK4) to motivate their (BK5c) $\sigma \simeq 1$ to get a 
$\left(
1 + \imath \partial_\tau \right)$ term --- because they wanted to cancel the
same factor from the \index{nonlinear}nonlinear term multiplying
$B(\vec{r}_\bot,\xi,\tau ; A)$.  This can be contrasted with
Porras\index{Porras}\cite{Porras-1999pra}, who was not interested in the
\index{nonlinear}nonlinear case but wanted instead to cancel terms in the
``$\nabla_\bot^2$'' diffraction (or self-focusing) term.  In fact using Brabec
and Krausz\cite{Brabec-K-1997prl} eqn. (BK4) as a
midpoint on the way to Brabec and Krausz\cite{Brabec-K-1997prl} eqn. (BK6) 
is not the best path, even for Brabec
and Krausz\cite{Brabec-K-1997prl} -- their early
selection of a $\left( 1 + \imath \partial_\tau \right)$ prefactor for
$\partial_\xi$ is unnecessary.  I show this here, but first hide all the
$\Upsilon$ dependent terms in $T_\Upsilon$ (as they are usually zero anyway),
and the rest of the RHS terms in $T_{RHS}$ because they remain unchanged.  

After some sign changes caused by moving factors of $\imath$ from the 
denominator to the numerator, and by moving one inside a set of $()$ 
brackets, we get

\begin{eqnarray}
0
&=&
    \left( 1 \pm \imath \sigma' \partial_\tau \right)
     q \partial_\xi
  A(\vec{r}_\bot,\xi,\tau) 
+
      \left( 
        \sigma \mp \sigma'
      \right) 
      \partial_\tau
  A(\vec{r}_\bot,\xi,\tau) 
-
    \left( 1 + \imath \sigma \partial_\tau \right)
    \left( 
    - \frac{ \alpha_0}{\beta_0}
    + \imath  \hat{D}'
    \right)
  A(\vec{r}_\bot,\xi,\tau) 
\nonumber
\\
&~&
 - 
    \frac{\imath }{2 \beta_0^2}
      \nabla_\bot^2 
    A(\vec{r}_\bot,\xi,\tau) 
  -
  \frac{2 \imath \pi }{n_0^2}
  \left[
        1 
      \pm  \imath \partial_\tau 
      \opm \imath g \sigma' \partial_\tau
   \omp
    g
    \left(
        1 
      - 
        \imath q \partial_\xi
    \right)
  \right]
  \left( 
         1 \pm \imath \partial_\tau  
  \right)
    B(\vec{r}_\bot,\xi,\tau ; A) 
  - 
    T_\Upsilon 
  - 
    T_{RHS}
,
\end{eqnarray}

\begin{eqnarray}
T_\Upsilon 
&=& 
  -
  \sigma   
  \Upsilon^\mp
  \left[
      \imath
    -  
      \left(
          \frac{ \alpha_0}{\beta_0}
        - \imath  \hat{D}'
      \right)
    - 
      \frac{\imath \sigma}{2} \Upsilon^\mp
    + 
      \sigma \partial_\tau
  \right]
    A(\vec{r}_\bot,\xi,\tau) 
\\
T_{RHS} 
&=& 
    \frac{\imath}{2}
  \left[
  -
    q^2 \partial_\xi^2
  +
    \left(
      \sigma^2 - \sigma'^2
     \right)
     \partial_\tau^2
  +
    \left(
      \frac{ \alpha_0}{\beta_0}
    - \imath  \hat{D}'
    \right)^2
  \right]
    A(\vec{r}_\bot,\xi,\tau) 
.
\end{eqnarray}

Hence  

\begin{eqnarray}
     q \partial_\xi
  A(\vec{r}_\bot,\xi,\tau) 
&=&
 -
  \frac{ \sigma \mp \sigma'}
       { 1 \pm \imath \sigma' \partial_\tau}
  \partial_\tau A(\vec{r}_\bot,\xi,\tau) 
 +
  \frac{ 1  +  \imath \sigma \partial_\tau}
       { 1 \pm \imath \sigma' \partial_\tau}
    \left( 
    - \frac{ \alpha_0}{\beta_0}
    + \imath  \hat{D}'
    \right)
  A(\vec{r}_\bot,\xi,\tau) 
  + 
    \frac{\imath }
         {2 \beta_0^2 \left( 1 \pm \imath \sigma' \partial_\tau \right)}
      \nabla_\bot^2 
    A(\vec{r}_\bot,\xi,\tau) 
\nonumber
\\
&~&
  +
    \frac{2 \imath \pi }{n_0^2}
    \frac{\left(1 + \imath \partial_\tau \right)^2}
         {\left( 1 \pm \imath \sigma' \partial_\tau \right)}
    B(\vec{r}_\bot,\xi,\tau ; A) 
  + 
    \frac{ T_\Upsilon + T_{RHS} }
         { 1 \pm \imath \sigma' \partial_\tau }
\label{exact-BKP-extra}
.
\end{eqnarray}

Now to avoid overly complex equations I set $\sigma'=\pm \sigma$ (i.e. 
$\beta'_1 = \beta_1$), and a \index{carrier}carrier appropriate to the group
velocity by choosing the upper sign), forcing the $\tau$ scaling for the field
to match the material rather than (e.g.) another field.  Note that in 
the case of multiple field components with different group velocities, it 
may be necessary to have $\beta'_1 \neq \beta_1$ in order to keep all
of the co-moving frames aligned.

One final step now gives us a Generalised Few-Cycle Envelope equation for a
propagating pulse in a \index{nonlinear}nonlinear medium.  It is in the style
of Brabec and Krausz\index{Brabec and Krausz}\cite{Brabec-K-1997prl}, but
unlike that of Brabec and Krausz\index{Brabec and
Krausz}\cite{Brabec-K-1997prl} (and of
Porras\index{Porras}\cite{Porras-1999pra}), it has no approximations beyond
that from the starting point, the permittivity convolution, and the separation
of $A$ and $A^*$ propagation equations.  Remembering to expand the $T_{RHS}$
and $T_\Upsilon$ terms as necessary, with $g=0$, we have the 
{\em Generalised Few-Cycle
Envelope Approximation (GFEA) equation}:

\begin{eqnarray}
     q \partial_\xi
  A(\vec{r}_\bot,\xi,\tau) 
&=&
    \left( 
    - \frac{ \alpha_0}{\beta_0}
    + \imath  \hat{D}'
    \right)
  A(\vec{r}_\bot,\xi,\tau) 
  + 
    \frac{\imath }
         {2 \beta_0^2 \left( 1 + \imath \sigma \partial_\tau \right)}
      \nabla_\bot^2 
    A(\vec{r}_\bot,\xi,\tau) 
\nonumber
\\
&~&
  +
    \frac{2 \imath \pi }{n_0^2}
    \frac{\left(1 + \imath \partial_\tau \right)^2}
         {\left( 1 + \imath \sigma \partial_\tau \right)}
    B(\vec{r}_\bot,\xi,\tau ; A) 
  + 
    \frac{ T_\Upsilon + T_{RHS} }
         { 1 + \imath \sigma \partial_\tau }
\label{exact-BKP}
.
\end{eqnarray}

\noindent
NOTE: If $g\neq0$, 
replace one of the $\left(1 + \imath \partial_\tau \right)$ numerator 
terms multiplying
$B$ in eqn.(\ref{exact-BKP}) with $  \left[
        1 
      \pm  \imath \partial_\tau 
      \opm \imath g \sigma' \partial_\tau
   \omp
    g
    \left(
        1 
      - 
        \imath q \partial_\xi
    \right)
  \right]$.

\subsection{The nonlinear ``few cycle'' term}

We might prefer to handle the \index{nonlinear}nonlinear term by basing it on
$\left(1 + \imath \partial_\tau \right)$, for example if we intended to
neglect the diffraction term entirely (e.g. as in Brabec and
Krausz\index{Brabec and Krausz}\cite{Brabec-K-1997prl}).  By starting at eqn
(\ref{exact-BKP}) we see the possibility of a nice expansion in $\delta = 1 -
\sigma$ by simply replacing the entire $B$ prefactor term in eqn
(\ref{exact-BKP}) with:

\begin{eqnarray}
    \frac{\left(1 + \imath \partial_\tau \right)^2}
         {\left( 1 + \imath \sigma \partial_\tau \right)}
    B(\vec{r}_\bot,\xi,\tau ; A) 
&=&
    \frac{\left(1 + \imath \partial_\tau \right)
          \left[
            \left(1 + \imath \sigma \partial_\tau \right)
           +\imath \left(1 - \sigma \right) \partial_\tau
          \right]
         }
         {\left( 1 + \imath \sigma \partial_\tau \right)}
    B(\vec{r}_\bot,\xi,\tau ; A) 
\\
&=&
    \left(1 + \imath \partial_\tau \right)
      \left[
        1
      +
        \frac{\imath \left(1 - \sigma \right) \partial_\tau}
             {\left(1 + \imath \sigma \partial_\tau \right)^2}
      \right]
    B(\vec{r}_\bot,\xi,\tau ; A) 
.
\label{BprefactorBK}
\end{eqnarray}

We could instead rearrange the $B$ term by aiming at the  $1 + \imath \sigma
\partial_\tau$ form of Porras\index{Porras}\cite{Porras-1999pra}:

\begin{eqnarray}
    \frac{\left(1 + \imath \partial_\tau \right)^2}
         {\left( 1 + \imath \sigma \partial_\tau \right)}
    B(\vec{r}_\bot,\xi,\tau ; A) 
&=&
    \frac{\left[1 + \imath \sigma \partial_\tau 
                  + \imath \left(1-\sigma\right) \partial_\tau
         \right]^2
         }
         {\left( 1 + \imath \sigma \partial_\tau \right)}
    B(\vec{r}_\bot,\xi,\tau ; A) 
\\
&=&
    \frac{ 
          \left(1 + \imath \sigma \partial_\tau \right)^2
           ~+~
           2 \left(1 + \imath \sigma \partial_\tau \right)
             \imath \left(1-\sigma\right) \partial_\tau
           ~-~
           \left(1-\sigma\right)^2 \partial_\tau
         }
         {\left( 1 + \imath \sigma \partial_\tau \right)}
    B(\vec{r}_\bot,\xi,\tau ; A) 
\\
&=&
    \left[
       \left(1 + \imath \sigma \partial_\tau \right)
     +
    \frac{ 2 \imath \left(1-\sigma\right) \partial_\tau
           + 
           \left(
             - 2 \sigma + 2 \sigma^2
             - 1 + 2 \sigma - \sigma^2
           \right)
           \partial_\tau^2
         }
         {\left( 1 + \imath \sigma \partial_\tau \right)}
    \right]
    B(\vec{r}_\bot,\xi,\tau ; A) 
\\
&=&
    \left(1 + \imath \sigma \partial_\tau \right)
    \left[
       1
     +
       \left(1-\sigma\right)
       \frac{ 2 \imath \partial_\tau
           + 
           \left(1+\sigma\right)
           \partial_\tau^2
           }
           {\left( 1 + \imath \sigma \partial_\tau \right)^2}
    \right]
    B(\vec{r}_\bot,\xi,\tau ; A) 
\label{BprefactorPorras}
\end{eqnarray}

It would be better, of course, to expand the few cycle term to a fixed
order explicitly -- there will be many expansions like the above, 
that when truncated to first order, are correct to within terms of
second order; but which differ from each other by amounts that 
are also of second order.  So:

\begin{eqnarray}
    \frac{\left(1 + \imath \partial_\tau   \right)^2}
         {\left( 1 + \imath \sigma \partial_\tau  \right)}
&=&
\left(1 + 2 \imath \partial_\tau  -  \partial_\tau^2 \right)
\times 
\left( 1 + \imath \sigma \partial_\tau  \right)^{-1}
\\
&=&
\left(1 + 2 \imath \partial_\tau  -  \partial_\tau^2  \right)
\nonumber
\\
&& ~~~~ 
  \times 
  \left[
         1 
       + \frac{-1 }{1!} 
         \imath \sigma \partial_\tau  
       + \frac{-1 \times -2 }{2!} 
         \imath^2 \sigma^2 \partial_\tau^2 
       + \frac{-1 \times -2 \times -3}{3!} 
         \imath^3 \sigma^3 \partial_\tau^3 
       + \mathscr{O}(\partial_\tau^4 )
  \right]
\\
&=&
\left(1 + 2 \imath \partial_\tau  -  \partial_\tau^2  \right)
\times 
  \left[ 
         1 
       - \imath \sigma \partial_\tau  
       - \sigma^2 \partial_\tau^2 
       + \imath \sigma^3 \partial_\tau^3 
       + \mathscr{O}(\partial_\tau^4 )
  \right]
\\
&=&
    1 
       + 2 \imath \partial_\tau  
       -  \partial_\tau^2 ~~~
   -  \imath \sigma \partial_\tau  
       - 2 \imath^2 \sigma \partial_\tau^2 
         +  \imath \sigma \partial_\tau^3 ~~~
   - \sigma^2 \partial_\tau^2 
       - 2 \imath \sigma^2 \partial_\tau^3 ~~~
   + \imath \sigma^3 \partial_\tau^3 ~~~
       + \mathscr{O}(\partial_\tau^4 )
\\
&=&
    1 
       + \imath 
           \left( 2 - \sigma \right)
           \partial_\tau  
       -  \left( 1 - 2 \sigma + \sigma^2 \right)
           \partial_\tau^2 
       +  \left( \imath \sigma - 2 \imath \sigma^2 + \imath \sigma^3 \right)
           \partial_\tau^3
       + \mathscr{O}(\partial_\tau^4 )
\\
&=&
    1 
       + \imath 
           \left( 2 - \sigma \right)
           \partial_\tau  
       -  \left( 1 - \sigma \right)^2
           \partial_\tau^2 
       +  \imath \sigma 
          \left( 1 - \sigma \right)^2
           \partial_\tau^3
       + \mathscr{O}(\partial_\tau^4 )
\label{eqn-expandB2nd}
.
\end{eqnarray}

It is interesting that this sytematic expansion gives yet another first order
correction to the non linear polarization -- but of course it only differs
from Brabec and Krausz's \cite{Brabec-K-1997prl} and Porras's
\cite{Porras-1999pra} by terms like $\left( 1-\sigma \right) \partial_\tau$,
which are second order corrections.  
Note again that Brabec and Krausz work in the case $\sigma=1$, whereas
Porras allows $\sigma \neq1 $, but does not consider nonlinear processes.





\section{Approximations: SEWA, SEEA, and GFEA}

The full generalised few-cycle equation (\ref{exact-BKP}) has a rather
complicated prefactor for the polarization term, and also the ``extra''
$T_{RHS}$ term.  If we want to make approximations that reduce it to Brabec
and Krausz's {\bf SEWA} (slowly evolving wave approximation),
Porras's\index{Porras}\cite{Porras-1999pra} {\bf SEEA} (slowly evolving
\index{envelope}envelope approximation), or some other form then we need to
consider these terms in detail, and understand in what limits these terms
might be simplified or neglected. Note that setting $T_\Upsilon=0$ is a matter
of chosen convention, and is not an approximation of any kind.

For example, Brabec and Krausz\index{Brabec and Krausz}\cite{Brabec-K-1997prl}
have a discussion in their PRL on p3284, after their eqn.(8) about the various
criteria that need to hold for their SEWA to be valid, all of which are that
various quantities must be slowly varying as the pulse propagates along $\xi$.
It is also instructive to see what parameter values or what terms are
neglected to see how my generalised equation (\ref{exact-BKP}) reduces to
Brabec and Krausz's {\bf SEWA} (slowly evolving wave approximation),
Porras\index{Porras}\cite{Porras-1999pra}'s {\bf SEEA} (slowly evolving
\index{envelope}envelope approximation), or to my {\bf GFEA} (generalised 
few-cycle \index{envelope}envelope approximation), which is in some sense 
equivalent to a ``best
of'' combination of the two others.

\begin{description}

\item[1. SEWA:]  Brabec and Krausz\index{Brabec and
Krausz}\cite{Brabec-K-1997prl} eqn.(6). In eqn. (\ref{exact-BKP}), set
$\sigma=1$, use $T_\Upsilon=0$ (since $\omega_\epsilon=\omega_0$)and ignore
$T_{RHS}$: 

\begin{eqnarray}
     q \partial_\xi
  A(\vec{r}_\bot,\xi,\tau) 
&=&
    \left( 
    - \frac{ \alpha_0}{\beta_0}
    + \imath  \hat{D}'
    \right)
  A(\vec{r}_\bot,\xi,\tau) 
  + 
    \frac{\imath }
         {2 \beta_0^2 \left( 1 + \imath \partial_\tau \right)}
      \nabla_\bot^2 
    A(\vec{r}_\bot,\xi,\tau) 
  +
    \frac{2 \imath \pi }{n_0^2}
    \left(1 + \imath \partial_\tau \right)
    B(\vec{r}_\bot,\xi,\tau ; A)
.
\label{eqn-BrabecKrausz}
\label{eqn-mySEWA}
\end{eqnarray}

\item[2. SEEA:] Porras\index{Porras}\cite{Porras-1999pra} eqn.(2).  In eqn. (\ref{exact-BKP}), set
$B(\vec{r}_\bot,\xi,\tau ; A)=0$, use $T_\Upsilon=0$ and ignore $T_{RHS}$:

\begin{eqnarray}
     q \partial_\xi
  A(\vec{r}_\bot,\xi,\tau) 
&=&
    \left( 
    - \frac{ \alpha_0}{\beta_0}
    + \imath  \hat{D}'
    \right)
  A(\vec{r}_\bot,\xi,\tau) 
  + 
    \frac{\imath }
         {2 \beta_0^2 \left( 1 + \imath \sigma \partial_\tau \right)}
      \nabla_\bot^2 
    A(\vec{r}_\bot,\xi,\tau) 
.
\label{eqn-Porras}
\label{eqn-mySEEA}
\end{eqnarray}

\item[3. GFEA:] Keeps the acuracies of both Brabec and Krausz\index{Brabec and Krausz}\cite{Brabec-K-1997prl} and Porras\index{Porras}\cite{Porras-1999pra}, 
whilst avoiding the more complicated parts of the full generalised eqn. 
(\ref{exact-BKP-extra}). 
In eqn. (\ref{exact-BKP}), set $1-\sigma \ll 1$ and use $T_\Upsilon=0$ and
ignore $T_{RHS}$:

\begin{eqnarray}
     q \partial_\xi
  A(\vec{r}_\bot,\xi,\tau) 
&=&
    \left( 
    - \frac{ \alpha_0}{\beta_0}
    + \imath  \hat{D}'
    \right)
  A(\vec{r}_\bot,\xi,\tau) 
  +
    \frac{2 \imath \pi }{n_0^2}
    \frac{  \left(1 + \imath \partial_\tau \right)^2 }
         {  \left(1 + \imath \sigma \partial_\tau \right) }
    B(\vec{r}_\bot,\xi,\tau ; A)
  + 
    \frac{\imath / \beta_0^2 }
         {2 \left( 1 + \imath \sigma \partial_\tau \right)}
      \nabla_\bot^2 
    A(\vec{r}_\bot,\xi,\tau) 
.
\label{eqn-GFEA}
\end{eqnarray}

\end{description}

The advantage of the generalised eqn. (\ref{exact-BKP}) over these is that I
could easily put $\sigma=1-\delta$ and do an expansion for small $\delta$ and
go beyond both Brabec and Krausz\index{Brabec and Krausz}\cite{Brabec-K-1997prl} and Porras\index{Porras}\cite{Porras-1999pra}.


The main point of complication with the generalised equation is the $T_{RHS}$
term, as it is second order in $\partial_\tau$ and $\partial_\xi$, as well as
having higher order terms in $\partial_\tau$ from the diffraction contribution
$\hat{D}'$.  These same terms (or simplified versions) were also
neglected by both Brabec and Krausz\index{Brabec and Krausz}\cite{Brabec-K-1997prl} and Porras\index{Porras}\cite{Porras-1999pra}.
Since the equation would be considerably easier to solve if
$T_{RHS}$ were negligible, I will examine it in carefully in order to see what
justification or constraints are required to do so. The term is:

\begin{eqnarray}
    \frac{ T_{RHS} }
         { 1 + \imath \sigma \partial_\tau }
&=&
  -
  \frac{\imath}{2}
  \left( 1 + \imath \sigma \partial_\tau \right)^{-1}
  \left[
    -
      \partial_\xi^2 
    +
        \left(
          \frac{ \alpha_0}{\beta_0}
        - 
          \imath \hat{D}'
        \right)^2
  \right]
  A(\vec{r}_\bot,\xi,\tau) 
.
\end{eqnarray}

Clearly this $T_{RHS}$ needs to be small compared to the
other terms in eqn. (\ref{exact-BKP}) if it is to be neglected. If it happens
that it {\em is} small, then the other, non-negligible, terms sum to close to
$\partial_\xi A(\vec{r}_\bot,\xi,\tau)$.  So we can self-consistently ignore 
$T_{RHS}$ if the following condition holds:

\begin{eqnarray}
  \left|
  \left( 1 + \imath \sigma \partial_\tau \right)^{-1}
    \left[
      \partial_\xi^2 
    -
      \left(
          \frac{ \alpha_0}{\beta_0}
          - 
          \imath \hat{D}'
      \right)^2
    \right]
    A(\vec{r}_\bot,\xi,\tau) 
  \right|
\ll
  \left|
    \partial_\xi A(\vec{r}_\bot,\xi,\tau)
  \right|
.
\label{condition-TRHS}
\end{eqnarray}

A nice way of dealing with the presence of the $\partial_\tau$ terms is to
fourier transform into the freqency domain, where 

\begin{eqnarray}
\imath \partial_\tau \rightarrow \Omega
\label{eqn-def-Omega}
.
\end{eqnarray}

  This enables us to avoid speculation about the possible time derivatives of
$A$, and instead constrain its frequency components.  However, this assumes 
knowlege of the all-time behaviour of the terms under consideration, so when
being careful it might be better to use a time-windowed transform
or similar.  This could then give us constraints valid over some finite
timescale relevant to the dynamics, without having to deal directly with
instantaneous derivatives.  

The
condition (\ref{condition-TRHS}) can be broken into two parts, which are

\begin{eqnarray}
& &
  \left|
    \partial_\xi^2 A(\vec{r}_\bot,\xi,\tau) 
  \right|
\ll
  \left|
    \left( 1 + \imath \sigma \partial_\tau \right)
    \partial_\xi A(\vec{r}_\bot,\xi,\tau)
  \right|
\label{condition-d2xi}
,
\\
& &
  \left|
    \left(
          \frac{ \alpha_0}{\beta_0}
          - 
          \imath \hat{D}'
    \right)^2
  A(\vec{r}_\bot,\xi,\tau) 
  \right|
\ll
  \left|
    \partial_\xi A(\vec{r}_\bot,\xi,\tau)
  \right|
\label{condition-dispersion2}
.
\end{eqnarray}

The second of these I assume holds as a further consequence of the 
``first order'' 
dispersion condition (\ref{condition-dispersion}) \xxlabel{d0} below.

\begin{subsubsection}{Note on the use of moduli }

The constraints we are attempting to apply are that the RHS term(s)
  has negligible effect on the propagation compared to that of 
  the LHS term(s); 
 the specific mathematical expression of this comparison is up to us.
The situation is complicated by the fact that either side can be complex,
 and will likely have a different complex argument (i.e. phase).  
Clearly the {\em largest} number we can make with the (hopefully small)
 LHS is given by the modulus, so that will give us an appropriate
 value for the LHS.
What to do with the RHS is less obvious, 
 because we would instead (to be cautious) want to pick a smallest 
 reasonable value; 
 but (e.g.) picking the minimum value of either the real or imaginary
 part would miss the point:
 indeed if the term were real, 
 the smallest imaginary part would be zero; 
 thus leading the condition to always fail.
This leaves us with little choice (as far as I can see) but to 
 use the modulus again;
 which in any case this would be the typical physicist's approach.

Perhaps the best justification for applying the modulus to both 
 sides in the comparison follows from the fact that 
 a rotation in the complex plane applied equally to both terms
 should not affect the outcome.
Hence we can rotate both terms so that the (hopefully large) RHS term 
 becomes real-valued; now, since
 only the LHS might be complex,
 taking its modulus gives a useful upper bound on the significance
 of its contribution to the dynamics.
This is equivalent to just taking the moduli of both terms.

\end{subsubsection}

\subsection{Evolution: $\partial_\xi^2$ approximation}

I can now constrain the evolution of the pulse in $\xi$ by evaluating how to
ensure that the $\partial_\xi^2$ terms is negligible. Starting with
eqn.(\ref{condition-d2xi}), 

\begin{eqnarray}
& &
  \left|
    \partial_\xi^2 A(\vec{r}_\bot,\xi,\tau) 
  \right|
\ll
  \left|
    \left( 1 + \imath \sigma \partial_\tau \right)
    \partial_\xi A(\vec{r}_\bot,\xi,\tau)
  \right|
\\
&\Longrightarrow&
  \left|
    \partial_\xi^2 A(\vec{r}_\bot,\xi,\tau) 
  \right|
\ll
  \left|
    \left( 1 + \sigma \omega \right)
    \partial_\xi \tilde{A}(\vec{r}_\bot,\xi,\Omega)
  \right|
\\
&\longrightarrow&
  \left|
    \partial_\xi \tilde{A}(\vec{r}_\bot,\xi,\Omega)
  \right|
\ll
  \left|
    \left( 1 + \sigma \omega \right)
    \tilde{A}(\vec{r}_\bot,\xi,\Omega)
  \right|
\\
&\longrightarrow&
  \left|
    \partial_\xi \tilde{A}(\vec{r}_\bot,\xi,\Omega)
  \right|
\ll
  \left|
    \tilde{A}(\vec{r}_\bot,\xi,\Omega)
  \right|
\label{condition-d2xi-final}
,
\end{eqnarray}

because (i) I  assume I can cancel (hopefully with no side effects) a
$\partial_\xi$ derivative term from either side \xxlabel{d0}; and (ii) I
tighten the constraint somewhat by relying on $\sigma \sim 1$ ({\em not}
$\sigma \simeq 1$) and that only considering positive frequencies means that
$\Omega > 0$.  Eqn (\ref{condition-d2xi}) is therefore revealed as a condition
that the \index{envelope}envelope function only changes slightly when propagated over distances
$\xi \sim 1$ (i.e. $z \sim 1/\beta_0$), and as such will depend on other 
terms in the evolution equation.

More carefully,
 if we assume we know the $\xi$ behaviour of the envelope $A$, 
 can use it to see how these constraints might hold in 
 spatial-frequency ($\kappa$) space 
 (since $\xi \leftrightarrow -\imath \kappa$), 
~
\begin{eqnarray}
& &
  \left|
    \partial_\xi^2 A(\vec{r}_\bot,\xi,\tau) 
  \right|
\ll
  \left|
    \left( 1 + \imath \sigma \partial_\tau \right)
    \partial_\xi A(\vec{r}_\bot,\kappa,\tau)
  \right|
\\
&\Longrightarrow&
  \left|
    \imath^2 \kappa^2 \tilde{A}(\vec{r}_\bot,\kappa,\tau) 
  \right|
\ll
  \left|
    - \imath \kappa 
    \left( 1 + \imath \sigma \partial_\tau \right)
    \tilde{A}(\vec{r}_\bot,\kappa,\tau)
  \right|
\\
\textrm{cancelling $\imath$ and $\kappa$ gives} 
~~~~ ~~~~
&\Longrightarrow&
  \left|
   \imath \kappa \tilde{A}(\vec{r}_\bot,\kappa,\tau) 
  \right|
\ll
  \left|
    -
    \left( 1 + \imath \sigma \partial_\tau \right)
    \tilde{A}(\vec{r}_\bot,\kappa,\tau)
  \right|
\\
\textrm{using moduli} 
~~~~ ~~~~
&\Longrightarrow&
  \left|
   \kappa \tilde{A}(\vec{r}_\bot,\kappa,\tau) 
  \right|
\ll
  \left|
    \left( 1 + \imath \sigma \partial_\tau \right)
    \tilde{A}(\vec{r}_\bot,\kappa,\tau)
  \right|
\end{eqnarray}

This gives us an expression rather like that as if the propagation 
 gave us different wavelength (wavevector) to that specified 
 by the carrier exponential.
We want the envelope to propagate with its spatially ($\xi$)
 behaviour to be peaked around small values of $\kappa$, 
 so that the bulk of the spatial variation is included in the 
 carrier exponential."

Note that the constraint will {\em always} be violated somewhere if the 
 spatial bandwidth of the propagation is too large (even if it has an
 e.g. exponential fall off).  The approximation therefore amounts 
 to ignoring such violations of the constraint in the (spatial-frequency)
 wings of the propagation, on the basis they are ``negligible'' 
 (which indeed seems perfectly reasonable).
In a simulation, 
 the LHS and RHS can be calculated as it progresses, 
 and the frequency counted, 
 and possible significance assessed (I have done this in OPA simulations).

\noindent
\xxref{d0} There may be complications I do not see, because these 
terms ($\hat{D}'$, etc)
do contain derivative terms.  However, since SVEA treatments ignore 
such issues, I do also.


\begin{center}
--- --- ---
\end{center}

The spatial evolution sub-condition (\ref{condition-d2xi-final}), the
dispersion sub-condition (\ref{condition-dispersion}) (c.f. condition
(\ref{condition-dispersion2})); and hence the {\em total} condition
(\ref{condition-TRHS}) will only hold if the other non-negligible terms in
eqn.(\ref{exact-BKP}) are similarly small compared to $\left|
A(\vec{r}_\bot,\xi,\tau) \right|$, viz:

\begin{eqnarray}
  \left|
    \left( 
      \frac{ \alpha_0}{\beta_0}
    - \imath  \hat{D}'
    \right)
  A(\vec{r}_\bot,\xi,\tau) 
  \right|
&\ll& 
  \left|
     A(\vec{r}_\bot,\xi,\tau)
  \right|
\label{condition-dispersion}
\\
  \left|
    \frac{\imath }
         {2 \beta_0^2 \left( 1 + \imath \sigma \partial_\tau \right)}
      \nabla_\bot^2 
    A(\vec{r}_\bot,\xi,\tau) 
  \right|
&\ll& 
  \left|
     A(\vec{r}_\bot,\xi,\tau)
  \right|
\label{condition-diffraction}
\\
  \left|
    \frac{2 \imath \pi }{n_0^2}
    \frac{  \left(1 + \imath \partial_\tau \right)^2 }
         {  \left(1 + \imath \sigma \partial_\tau \right) }
    B(\vec{r}_\bot,\xi,\tau ; A)
  \right|
&\ll& 
  \left|
     A(\vec{r}_\bot,\xi,\tau)
  \right|
\label{condition-nonlinear}
.
\end{eqnarray}


Note that we could use the alternative eqn (\ref{BprefactorBK}) in
condition (\ref{condition-nonlinear}).

I will assume condition (\ref{condition-dispersion}) implies that
(\ref{condition-dispersion2}) also holds \xxlabel{d0}, leaving us with four
conditions in total (\ref{condition-d2xi-final}, \ref{condition-dispersion},
\ref{condition-diffraction}, \ref{condition-nonlinear}).  Note that
Brabec and Krausz\index{Brabec and Krausz}\cite{Brabec-K-1997prl} split (\ref{condition-dispersion}) into multiple pieces, which
will be discussed in the following subsection.

\subsection{Dispersion: $\partial_\tau$ Approximation}

I will now treat the dispersion condition (\ref{condition-dispersion}) in the
above approximations and associated conditions.  Note that Brabec and Krausz\index{Brabec and Krausz}\cite{Brabec-K-1997prl} claim
that their SEWA ``does not explicitly impose a limitation on the pulse width'';
however this is rather misleading as shortly afterward they introduce a pulse
duration $\tau_p$ which is used in the inequalities constraining the material
parameters, which then give the region of validity of the SEWA.  Even the 
weakest
statement we might make about $\tau_p$ needs to state that it {\em does}
constrain the SEWA, because it (further) constrains the material parameters! 
For a given set of material parameters, there will be some pulse width
limitation, although it might well be the few-cycles we hope to describe.

I now break up eqn. (\ref{condition-dispersion}) into parts containing single
factors of $\gamma_m$.  Note that I need to exclude any terms including
$\beta_0$ and $\beta_1$ in $\gamma_m$ as  they were treated separately in the 
analysis. I thus write $\gamma'_m=\gamma_m$ for
$m\geq2$, and $\gamma'_0=\alpha_0$, $\gamma_1'=\alpha_1$.  The inequality is
then

\begin{eqnarray}
  \left|
    \left( 
      \frac{ \omega_0^m \gamma'_m}{\beta_0 m!}
      \imath^m
      \partial_\tau^m
    \right)
  A(\vec{r}_\bot,\xi,\tau) 
  \right|
&\ll& 
  \left|
     A(\vec{r}_\bot,\xi,\tau)
  \right|
\label{condition-gammaAt}
\\
  \left|
    \left( 
      \frac{ \omega_0^m \gamma'_m}{\beta_0 m!}
      \Omega^m
    \right)
  \tilde{A}(\vec{r}_\bot,\xi,\Omega)
  \right|
&\ll& 
  \left|
     \tilde{A}(\vec{r}_\bot,\xi,\Omega)
  \right|
,
\label{condition-gammaA}
\end{eqnarray}

where the second line has been fourier transformed in time.  Note that while 
both $\omega_0$ and $\gamma'$ have units, $\Omega$ does not, as it is the 
counterpart of the dimensionless (scaled) $\tau$ (see (\ref{eqn-def-Omega})).

\noindent
{\bf To Do:} 
Now get from $\Omega$ to a $\tau_\delta$ or Brabec and Krausz\index{Brabec and Krausz}\cite{Brabec-K-1997prl} $\tau_p$, by 
some physical motivation justifying $\tau_\delta = 2\pi / \Omega$.

I could treat the time-domain condition (\ref{condition-gammaAt}) qualitatively
as follows:  introduce $\tau_\delta \simeq A / \left| \partial_\tau A \right|$,
the time for which a rate of change of $\partial_\tau A$ would accumulate (in
absolute value) to A; and is thus some measure of how much time it takes the
\index{envelope}envelope to change significantly, and is thus hopefully 
something we can relate
to the  pulse width (or at least the width of one ``bump'' on the pulse
envelope).  Also, we assume that the higher derivatives of A are related, with
$\left| \partial_\tau^m A / m! \right| / A \simeq \tau_\delta^{-m}$
\xxlabel{x1}.  This is essentially the same
parameter as Brabec and Krausz\index{Brabec and Krausz}\cite{Brabec-K-1997prl}'s $\tau_p$, but scaled into my dimensionless picture: 
$\tau_p \equiv \tau_\delta /
\omega_0$.  Using this $\tau_\delta$ we can rewrite eqn. 
(\ref{condition-gammaAt}) as

\begin{eqnarray}
  \left|
      \frac{ \omega_0^m \left|\gamma'_m\right|}{\beta_0}
      \tau_\delta^{-m}
  \right|
&\ll& 
1
\label{condition-gamma}
\\
      \frac{ \tau_\delta^m }
           { \omega_0^m \left|\gamma'_m\right| }
\equiv
      \frac{ \tau_p^m }
           { \left|\gamma'_m\right| }
&\gg&
      \beta_0^{-1} \sim L_{\gamma,m}
\end{eqnarray}

This final condition is the same as those for $\alpha_m$ and 
$\beta_m$ in Brabec and Krausz\index{Brabec and Krausz}\cite{Brabec-K-1997prl}, since ($\beta_0$ and wavelength comment),
and ($L_{\gamma,m}$ justification cf BK).

\noindent
\xxref{x1}: The factoring of the $m!$ into $\partial_\tau^m A$ seem the 
most mathematically most sensible thing to do by analogy to the terms 
in Taylor expansions, and expansions of exponential functions (etc).

\noindent
{\bf Note:} Is it possible to derive a true time-domain treatment of the 
constraints  by replacing the $\gamma_m$ by $\partial_\omega k(\omega)$, 
and relate it back to the time domain $\epsilon(t-t_0)$ (or moments thereof).

\noindent
{\bf Note:} the eqn. (\ref{condition-dispersion2}) $\hat{D}'^2$ terms.


\subsection{Diffraction: $\nabla_\bot^2$ approximations}

Treating the $\partial_\tau$ in condition (\ref{condition-diffraction}) by
fourier transform, as above, we can use the fact that for 
gaussian beams with a beam waist $w_0$,  we have $\nabla_\bot^2
A(\vec{r}_\bot,\xi,\tau) \sim w_0^{-2}$; similar statements could be made
for other typical beam profiles.  This leads to the diffraction
constraint on the SEWA becoming

\begin{eqnarray}
    \left(1 + \sigma \Omega \right)
\beta_0^2 w_0^2 
&\gg& 
  1
.
\label{condition-beamwaist}
\end{eqnarray}

Comparing this to the comparable condition in Brabec and Krausz\index{Brabec and Krausz}\cite{Brabec-K-1997prl} after their eqn (8)
(i.e. in my units $\beta_0^2 w_0^2 \gg 1 $), we see that they are the same
except for the $\Omega$ term, so that my condition is in fact somewhat less
restrictive than theirs.  This is because considering positive frequencies only
means that $\Omega > 0$; and $\Omega \sim 1$ for $\partial_\tau$ modulations 
much less than the variation of the carrier frequency.


\subsection{Nonlinearity: $B(\vec{r}_\bot,\xi,\tau ; A)$ approximations}

The \index{nonlinear}nonlinearity constraint (\ref{condition-nonlinear}) is very complicated,
so I ignore the term in square brackets $\left[ ... \right]$.  Then, 
treating the $\partial_\tau$ in condition (\ref{condition-nonlinear})
by fourier transform, as above, and
using $\sigma \simeq 1$, the \index{nonlinear}nonlinearity constraint on the SEWA becomes 

\begin{eqnarray}
  \frac{2\pi}{n_0^2}
  \frac{ \left(1 + \sigma \Omega \right) }
       { \left(1 + \Omega \right)^2 }
&\gg& 
  \frac
  {
  \left| 
    \tilde{B}(\vec{r}_\bot,\xi,\Omega ; A)
  \right|
  }
  { \left|
     \tilde{A}(\vec{r}_\bot,\xi,\Omega)
    \right|
  }
.
\label{condition-nonlinearB}
\end{eqnarray}

Comparing this to the comparable condition in Brabec and Krausz\index{Brabec
and Krausz}\cite{Brabec-K-1997prl} after their eqn (BK8), we see that the same
comments as for the diffraction hold -- my condition is somewhat less
restrictive than theirs.  Of course it may be convenient to simplify the LHS
of eqn. (\ref{condition-nonlinearB}) with various small $\Omega$ expansions.

In both cases I could instead include the $\partial_\tau$ corrections
qualitatively by replacing $\partial_\tau \rightarrow \tau_\delta^{-1}$ before
proceeding from (\ref{condition-diffraction}) and (\ref{condition-nonlinear});
but even with $\tau_\delta \simeq 1$ this will not alter the contraints 
greatly.




\printindex


\begin{thebibliography}{14}
\expandafter\ifx\csname natexlab\endcsname\relax\def\natexlab#1{#1}\fi
\expandafter\ifx\csname bibnamefont\endcsname\relax
  \def\bibnamefont#1{#1}\fi
\expandafter\ifx\csname bibfnamefont\endcsname\relax
  \def\bibfnamefont#1{#1}\fi
\expandafter\ifx\csname citenamefont\endcsname\relax
  \def\citenamefont#1{#1}\fi
\expandafter\ifx\csname url\endcsname\relax
  \def\url#1{\texttt{#1}}\fi
\expandafter\ifx\csname urlprefix\endcsname\relax\def\urlprefix{URL }\fi
\providecommand{\bibinfo}[2]{#2}
\providecommand{\eprint}[2][]{\url{#2}}

\bibitem[{\citenamefont{Kinsler and New}(2003)}]{Kinsler-N-2003pra}
\bibinfo{author}{\bibfnamefont{P.}~\bibnamefont{Kinsler}} \bibnamefont{and}
  \bibinfo{author}{\bibfnamefont{G.~H.~C.} \bibnamefont{New}},
  \bibinfo{journal}{Phys. Rev. A} \textbf{\bibinfo{volume}{67}},
  \bibinfo{pages}{023813} (\bibinfo{year}{2003}),
  \urlprefix\url{http://link.aps.org/abstract/PRA/v67/e023813}.

\bibitem[{\citenamefont{Kinsler and New}(2002)}]{Kinsler-N-2002longFCOPO}
\bibinfo{author}{\bibfnamefont{P.}~\bibnamefont{Kinsler}} \bibnamefont{and}
  \bibinfo{author}{\bibfnamefont{G.~H.~C.} \bibnamefont{New}},
  \bibinfo{journal}{arXiv.org} \textbf{\bibinfo{volume}{physics}},
  \bibinfo{pages}{0212016} (\bibinfo{year}{2002}),
  \urlprefix\url{http://arXiv.org/physics/0212016}.

\bibitem[{\citenamefont{Brabec and Krausz}(1997)}]{Brabec-K-1997prl}
\bibinfo{author}{\bibfnamefont{T.}~\bibnamefont{Brabec}} \bibnamefont{and}
  \bibinfo{author}{\bibfnamefont{F.}~\bibnamefont{Krausz}},
  \bibinfo{journal}{Phys. Rev. Lett.} \textbf{\bibinfo{volume}{78}},
  \bibinfo{pages}{3282} (\bibinfo{year}{1997}),
  \urlprefix\url{http://link.aps.org/abstract/PRL/v78/p3282}.

\bibitem[{\citenamefont{Porras}(1999)}]{Porras-1999pra}
\bibinfo{author}{\bibfnamefont{M.~A.} \bibnamefont{Porras}},
  \bibinfo{journal}{Phys. Rev. A} \textbf{\bibinfo{volume}{60}},
  \bibinfo{pages}{5069} (\bibinfo{year}{1999}),
  \urlprefix\url{http://link.aps.org/abstract/PRA/v60/p5069}.

\bibitem[{\citenamefont{Fleck}(1970)}]{Fleck-1970prb}
\bibinfo{author}{\bibfnamefont{J.~A.} \bibnamefont{Fleck}},
  \bibinfo{journal}{Phys. Rev. B} \textbf{\bibinfo{volume}{1}},
  \bibinfo{pages}{84} (\bibinfo{year}{1970}),
  \urlprefix\url{http://link.aps.org/abstract/PRB/v1/p84}.

\bibitem[{\citenamefont{G.L.~Lamb}(1971)}]{Lamb-1971rmp}
\bibinfo{author}{\bibfnamefont{J.}~\bibnamefont{G.L.~Lamb}},
  \bibinfo{journal}{Rev. Mod. Phys.} \textbf{\bibinfo{volume}{43}}
  (\bibinfo{year}{1971}),
  \urlprefix\url{http://link.aps.org/abstract/RMP/v43/p99}.

\bibitem[{\citenamefont{Kanetsyan}(2002)}]{Kanetsyan-2002iqec}
\bibinfo{author}{\bibfnamefont{E.~G.} \bibnamefont{Kanetsyan}},
  \bibinfo{journal}{IQEC Techn. Digest} p. \bibinfo{pages}{465}
  (\bibinfo{year}{2002}), \bibinfo{note}{abstract number QThL20}.

\bibitem[{\citenamefont{Xiao et~al.}(2002)\citenamefont{Xiao, Wang, and
  Xu}}]{Xiao-WX-2002pra}
\bibinfo{author}{\bibfnamefont{J.}~\bibnamefont{Xiao}},
  \bibinfo{author}{\bibfnamefont{Z.}~\bibnamefont{Wang}}, \bibnamefont{and}
  \bibinfo{author}{\bibfnamefont{Z.}~\bibnamefont{Xu}}, \bibinfo{journal}{Phys.
  Rev. A} \textbf{\bibinfo{volume}{65}}, \bibinfo{pages}{031402(R)}
  (\bibinfo{year}{2002}),
  \urlprefix\url{http://link.aps.org/abstract/PRA/v65/e031402}.

\bibitem[{\citenamefont{Kinsler}(2002)}]{Kinsler-TLAFC}
\bibinfo{author}{\bibfnamefont{P.}~\bibnamefont{Kinsler}},
  \emph{\bibinfo{title}{Two level atoms and the few-cycle regime}}
  (\bibinfo{publisher}{Personal Report}, \bibinfo{year}{2002}),
  \urlprefix\url{file:twolevelatom.dvi}.

\bibitem[{\citenamefont{Chelkowski and Bandrauk}(2002)}]{Chelkowski-B-2002pra}
\bibinfo{author}{\bibfnamefont{S.}~\bibnamefont{Chelkowski}} \bibnamefont{and}
  \bibinfo{author}{\bibfnamefont{A.~D.} \bibnamefont{Bandrauk}},
  \bibinfo{journal}{Phys. Rev. A} \textbf{\bibinfo{volume}{65}},
  \bibinfo{pages}{061802(R)} (\bibinfo{year}{2002}),
  \urlprefix\url{http://link.aps.org/abstract/PRA/v65/e061802}.

\bibitem[{\citenamefont{Shen}(1984)}]{Shen-PNLO}
\bibinfo{author}{\bibfnamefont{Y.~R.} \bibnamefont{Shen}},
  \emph{\bibinfo{title}{Principles of nonlinear optics}}
  (\bibinfo{publisher}{Wiley}, \bibinfo{year}{1984}).

\bibitem[{\citenamefont{Kinsler et~al.}(2005)\citenamefont{Kinsler, Radnor, and
  New}}]{Kinsler-RN-2005pra}
\bibinfo{author}{\bibfnamefont{P.}~\bibnamefont{Kinsler}},
  \bibinfo{author}{\bibfnamefont{S.~B.~P.} \bibnamefont{Radnor}},
  \bibnamefont{and} \bibinfo{author}{\bibfnamefont{G.~H.~C.}
  \bibnamefont{New}}, \bibinfo{journal}{Phys. Rev. A}
  \textbf{\bibinfo{volume}{72}}, \bibinfo{pages}{063807}
  (\bibinfo{year}{2005}).

\bibitem[{\citenamefont{Kinsler}(2006)}]{Kinsler-FLECK}
\bibinfo{author}{\bibfnamefont{P.}~\bibnamefont{Kinsler}},
  \bibinfo{journal}{arXiv.org} \textbf{\bibinfo{volume}{physics}},
  \bibinfo{pages}{0611216} (\bibinfo{year}{2006}),
  \urlprefix\url{http://arxiv.org/abs/physics/0611216}.

\bibitem[{\citenamefont{Rothenburg}(1992)}]{Rothenburg-1992ol}
\bibinfo{author}{\bibfnamefont{J.~E.} \bibnamefont{Rothenburg}},
  \bibinfo{journal}{Opt. Lett.} \textbf{\bibinfo{volume}{17}},
  \bibinfo{pages}{1340} (\bibinfo{year}{1992}), \bibinfo{note}{referred to by
  Porras[2], and Brabec-Krausz[15].},
  \urlprefix\url{http://ol.osa.org/abstract.cfm?id=11337}.

\end{thebibliography}
\end{document}